\documentclass[a4paper,11pt]{article}
\pdfoutput=1 

\usepackage{jheppub} 

\usepackage[T1]{fontenc} 
\usepackage{comment}
\usepackage{tikz}
\usepackage{amsmath,amssymb}

\title{3d conformal fields with manifest $sl(2,\mathbb{C})$}

\author{Dmitry Ponomarev}

\affiliation{Institute for Theoretical and Mathematical Physics,\\
Lomonosov Moscow State University, Lomonosovsky avenue, Moscow, 119991, Russia}
\affiliation{I.E. Tamm Theory Department, Lebedev Physical Institute,\\
 Leninsky avenue, Moscow, 119991, Russia}

\emailAdd{ponomarev@lpi.ru}

\abstract{In the present paper we construct all short representation of $so(3,2)$ with the $sl(2,\mathbb{C})$ symmetry made manifest due to the use of $sl(2,\mathbb{C})$ spinors. This construction has a natural connection to the spinor-helicity formalism for massless fields in AdS${}_4$ suggested earlier. We then study unitarity of the resulting representations, identify them as the lowest-weight modules and as conformal fields in the three-dimensional Minkowski space. Finally, we compare these results with the existing literature and discuss the properties of these  representations under contraction of $so(3,2)$ to the Poincare algebra.}

\begin{document} 
\maketitle
\flushbottom

\section{Introduction}

Construction of interacting theories of massless higher spin fields is a promising, but at the same time a very challenging problem of modern physics. For a long time it is known that under some reasonable assumptions -- primarily, locality in one or another form -- massless higher spin fields cannot interact in flat space, see e.g. \cite{Weinberg:1964ew,Coleman:1967ad} and \cite{Bekaert:2010hw} for review. To circumvent these no-go results it was suggested that non-trivial higher spin theories can exist in (A)dS \cite{Fradkin:1987ks}. Significant progress in this direction was made over the years and, in particular, concrete  higher-spin theories were proposed \cite{Vasiliev:1990en,Vasiliev:2003ev}. Though, currently, locality of these theories is a subject of active research \cite{Giombi:2009wh,Boulanger:2015ova,Vasiliev:2016xui,Didenko:2020bxd}, independent support for consistency of higher-spin theories in AdS came from the AdS/CFT correspondence \cite{Sezgin:2002rt,Klebanov:2002ja}. 
Namely, the conjecture states that higher-spin theories in AdS are dual to a class of simple conformal theories, among which is the free $O(N)$ vector model. This allows one if not to entirely define, but at least to access significant amount of information about higher spin theories holographically.
In particular, holography was used recently to derive lower-order vertices and analyze locality of the bulk theory \cite{Bekaert:2014cea,Bekaert:2015tva,Sleight:2016dba,Sleight:2017pcz,Ponomarev:2017qab}.

Along with that, some progress was achieved in understanding higher spin theories in flat space using the light-cone gauge approach. Based on earlier works \cite{Metsaev:1991mt,Metsaev:1991nb} in \cite{Ponomarev:2016lrm} chiral higher spin theories were proposed\footnote{For a related earlier result, see \cite{Devchand:1996gv}.}. These theories are manifestly local, contain only cubic vertices and, at the same time, are consistent to all orders in interactions. It should be emphasised, that the action of chiral theories is not real in the Lorentzian signature, so these theories do not yet provide a desired solution to the higher spin problem in flat space. Instead, these should be regarded as natural higher spin counterparts of self-dual Yang-Mills and self-dual gravity theories -- in fact, this connection can be made very precise \cite{Ponomarev:2017nrr} -- and as the latter theories, they require a parity-invariant completion. While this completion is obstructed by non-localities and as of now there is no clear understanding whether locality can be relaxed without bringing pathologies into the theory and making the program of perturbative construction of higher spin interactions ill-defined, simplicity of chiral higher spin theories and their uniformity with the lower spin counterparts is encouraging. For recent progress on chiral higher-spin theories, see \cite{Skvortsov:2018jea,Skvortsov:2020wtf,Skvortsov:2020gpn,Skvortsov:2020pnk}.

Another intriguing feature of chiral higher spin theories is that these involve vertices, that cannot be naturally written in terms of tensor fields and their derivatives contracted in a manifestly covariant manner \cite{Bengtsson:2014qza,Conde:2016izb} -- the approach, that we will refer to as the manifestly covariant one. Exotic light-cone vertices, which are lacking in the manifestly covariant description also have natural counterparts in the spinor-helicity formalism. In fact, there is a close connection between the spinor-helicity formalism and the light-cone gauge approach \cite{Ananth:2012un,Bengtsson:2016jfk,Ponomarev:2016cwi}.

Holographic higher spin theories in AdS and chiral theories, though, constructed from completely different considerations, seem to be intimately related. In particular, it was shown that, once the flat space limit is taken\footnote{The flat space limit of AdS cubic vertices for massless higher-spin fields was first discussed in \cite{Boulanger:2008tg}. In particular, it was pointed out that this limit can be made regular, thus, resulting in consistent flat space vertices.}, some cubic couplings of the former theories agree with those of the latter\footnote{In \cite{Skvortsov:2015pea} an observation was made that this is the case for $0-0-s$ vertices, while for cubic interactions with any spinning fields this was confirmed in \cite{Sleight:2016dba}.}. This agreement was checked for higher derivative vertices -- those, for which the flat-space limit is unambiguous and covariant formalism has access to\footnote{Recently, cubic vertices for massless fields in AdS${}_4$ were constructed in the light-cone formalism  \cite{Metsaev:2018xip}. This construction does involve additional vertices compared to those available in the covariant formalism. For the analysis at the next order, that allows to fix all cubic coupling constants, and for the discussion of the light-cone formalism in the holographic context, see  \cite{Skvortsov:2018uru}.}. Unless a mere coincidence, this matching may indicate  that chiral higher spin theories have some sort of a holographic description. If this is, indeed, the case, this can provide the first explicit example of flat space holography, which was mainly explored at a kinematic level  so far. For a far from complete list of references on recent advances in flat holography, see \cite{Bagchi:2016bcd,Pasterski:2017ylz,Schreiber:2017jsr,Ciambelli:2018wre,Donnay:2018neh,Stieberger:2018onx,Fan:2019emx,Adamo:2019ipt}.

 Whether any holographic construction underlies chiral higher spin theories or not, the manifestly covariant formalism is not suitable to deal with it as it does not allow to capture all relevant vertices in flat space. Therefore,  to proceed in this direction, one first needs to develop a framework that would be applicable to theories in AdS and, at the same time, would be able to deal with all interactions, relevant for the chiral theory. In the following we will focus on the extensions of the spinor-helicity formalism, as these seem to be simpler than analogous extensions of  the light-cone approach.

In the series of papers  the spinor-helicity formalism for massless fields in AdS${}_4$ was developed  and some simple amplitudes were explored \cite{Nagaraj:2018nxq,Nagaraj:2019zmk,Nagaraj:2020sji}\footnote{For other closely related approaches we refer the reader to \cite{Bolotin:1999fa,Maldacena:2011nz,Adamo:2012nn,Colombo:2012jx,Didenko:2012tv,Gelfond:2013xt,Skinner:2013xp,David:2019mos}.}. In the present paper we will focus on other important representations, relevant for the higher-spin holography -- Dirac singletons \cite{Dirac:1963ta}, which are better known as conformal scalar and spinor fields. It is worth cautioning the reader, that, while typically the spinor-helicity formalism refers to a more concrete set of techniques that allow to manipulate amplitudes of massless fields in four-dimensional flat space efficiently, in the context of short representations of $so(3,2)$, we will use this notion more broadly, rather, as a general idea of employing $sl(2,\mathbb{C})\simeq so(3,1)$ spinors to make Lorentz symmetry manifest.

The idea of employing $sl(2,\mathbb{C})$ spinors to construct representations of $so(3,2)$ in the higher spin context is not new. In fact, this is exactly how massless fields are represented in the Vasiliev theory. This approach has also been extended to massive fields of general spin and mass, which includes partially massless and continuous spin representations \cite{Skvortsov:2006at,Ponomarev:2010st,Khabarov:2019dvi}. In these cases, $so(3,2)$ modules naturally decompose into infinite sets of tensor representations of $sl(2,\mathbb{C})$ and the remaining generators -- deformed translations -- mix these $sl(2,\mathbb{C})$ representations with each other. For other works employing this or closely related formalisms, see \cite{Vasiliev:1992gr,Barabanshchikov:1996mc,Shaynkman:2000ts,Iazeolla:2008ix,Boulanger:2008up,deAzcarraga:2014hda,Boulanger:2014vya,Zinoviev:2015sra}. In particular, it first appeared in the context of unfolding of massive fields in AdS${}_3$ in \cite{Vasiliev:1992gr}, while various mathematical aspects of this approach were explored in \cite{Iazeolla:2008ix}.

What makes the singleton case conceptually different is that singletons are short representations, in the sense that unlike massive and massless AdS fields, whose degrees of freedom can be labelled with three continuous variables, the singleton degrees of freedom can be labelled with only two. For this reason, while the former can be naturally interpreted as on-shell fields in the bulk, the latter correspond to on-shell fields on the AdS boundary. 
Moreover, while the decomposition of $so(3,2)$ modules into tensor modules of the Lorentz algebra has a natural interpretation of the bulk derivative expansion, singletons under restriction to $sl(2,\mathbb{C})$ give irreducible representations \cite{Angelopoulos:1980wg,Angelopoulos:1997ij,Bekaert:2011js}, which are, clearly, infinite-dimensional.
  Instead of decomposing the latter into infinite series of tensor representations, which raises issues with convergence, we find it more natural to keep them as they are, that is in the form of infinite-dimensional $sl(2,\mathbb{C})$ modules. 

The presence of infinite-dimensional representations of $sl(2,\mathbb{C})$ brings few modifications into the standard procedure \cite{Skvortsov:2006at,Ponomarev:2010st,Khabarov:2019dvi} of constructing $so(3,2)$ modules with manifest Lorentz symmetry. It is the aim of the present paper to study them.
More precisely,  we will classify short $so(3,2)$ modules using this framework and compare the results with the existing literature. Our main focus will be on those representations of $so(3,2)$, that reduce  to a finite number of $sl(2,\mathbb{C})$ modules, which is the case relevant for short $so(3,2)$ modules or, equivalently, conformal fields on the boundary of AdS${}_4$. We will then explore their flat limit in a given form. 

This paper is organised as follows. In section \ref{sec:2} we review the standard facts on $sl(2,\mathbb{C})$ modules. We then explain the approach and present a general solution for  $so(3,2)$ modules in this framework. In section  \ref{sec:3} we study the pattern of module decomposition and classify all possible cases in which this decomposition gives rise to short modules. Next,  in section \ref{sec:4} we study their unitarity. In section \ref{sec:5} we find the lowest weight vectors and their eigenvalues, which then allows us to compare our results with the literature. In section \ref{sec:6} we discuss the flat space limit and then we present our conclusions.

\section{The setup}
\label{sec:2}

In this section we construct $so(3,2)$ modules with $sl(2,\mathbb{C})$ realized manifestly in terms of $sl(2,\mathbb{C})$ spinors. 
This problem was addressed before in the process of unfolding of massive higher-spin fields \cite{Ponomarev:2010st,Khabarov:2019dvi}, see also \cite{Misuna:2020fck} for related results. The major difference of these results with our analysis is that we are primarily interested in the case that involves infinite-dimensional representations of $sl(2,\mathbb{C})$. Infinite-dimensional representations of the Lorentz group first appeared in the field theoretic context in \cite{Majorana:1968zz}. These representations then were further studied in \cite{Dirac:1945cm,Gelfand:1947ui,Harish1947,Bargmann:1946me,Gelfand:1948ui}. For a
 comprehensive introduction to representations of $sl(2,\mathbb{C})$ we refer the reader to \cite{Gelfand} and for our conventions to appendix \ref{App:conv}. For recent discussions of infinite-dimensional representations of the Lorentz group in a field-theoretic context, see e. g. \cite{Fedoruk:1994ij,Bekaert:2009pt,Basile:2016aen}.

\subsection{Representations of $sl(2,\mathbb{C})$}

Our goal is to construct representations of the AdS${}_4$ isometry algebra  $so(3,2)$, by employing $sl(2,\mathbb{C})$ spinors and thus making the Lorentz symmetry $so(3,1)\simeq sl(2,\mathbb{C})$ manifest. The action of the Lorentz algebra can be realized as
\begin{equation}
\label{4may1}
{ J}_{\alpha\beta}=i\left(\lambda_{\alpha}\frac{\partial}{\partial \lambda^\beta}+ \lambda_{\beta}\frac{\partial}{\partial \lambda^\alpha} \right), \qquad \bar{ J}_{\dot\alpha\dot\beta}=i\left(\bar\lambda_{\dot\alpha}\frac{\partial}{\partial \bar\lambda^{\dot\beta}}+ \bar\lambda_{\dot\beta}\frac{\partial}{\partial \bar\lambda^{\dot\alpha}} \right),
\end{equation}
where the operators are assumed to act on functions $f(\lambda,\bar\lambda)$ on $\mathbb{C}^2/\{ 0\}$. Obviously, generators in (\ref{4may1}) commute with
\begin{equation}
\label{4may2x1}
\hat{\bar N}\equiv \bar\lambda^{\dot\alpha}\frac{\partial}{\partial \bar\lambda^{\dot\alpha}}, \qquad \hat N\equiv \lambda^\alpha \frac{\partial}{\partial \lambda^\alpha},
\end{equation}
so constraints
\begin{equation}
\label{4may3x1}
\hat{\bar N}f(\lambda,\bar\lambda) = {\bar N}f(\lambda,\bar\lambda), \qquad \hat Nf(\lambda,\bar\lambda) = Nf(\lambda,\bar\lambda) 
\end{equation}
with any complex $N$ and $\bar N$ are $sl(2,\mathbb{C})$ invariant. To ensure that $f$ is single valued, one has to require
\begin{equation}
\label{4may4x1}
\bar N-N \in \mathbb{Z}.
\end{equation}
Note that  $\bar N$  and $N$ are not necessarily complex conjugate to each other.

A representation of $sl(2,\mathbb{C})$ carried by functions that satisfy the above conditions will be denoted ${\cal V}^{N,\bar N}$. For genuine $N$ and $\bar N$, these representations are infinite-dimensional and irreducible. For specific values of $N$, $\bar N$ the structure of ${\cal V}^{N,\bar N}$ is more complicated. For example, when both $N$ and $\bar N$ are non-negative integers, ${\cal V}^{N,\bar N}$ admits an invariant subspace given by polynomials of the respective homogeneity degrees. For more comprehensive discussion of the reducibility structure of ${\cal V}^{N,\bar N}$ we refer the reader to \cite{Gelfand}. Below, we will be primarily interested in representations with genuine $N$ and $\bar N$.

Homogeneity constraints (\ref{4may3x1}) imply that for $f\in {\cal V}^{N,\bar N}$ one has 
\begin{equation}
\label{18feb2}
f(\alpha\lambda,\bar\alpha\bar\lambda)=\alpha^N\bar\alpha^{\bar N}f(\lambda,\bar\lambda),
\end{equation}
hence, to fully determine $f$, it is enough to define it for a single point of each complex line that passes trough the origin. This allows one to identify $ {\cal V}^{N,\bar N}$ with functions on the Riemann sphere, $\mathbb{CP}^1$. In the following, we will sometimes need to get rid off redundant dependence in $f$ due to homogeneity constraints. To achieve that we will represent $ {\cal V}^{N,\bar N}$ by functions on $\lambda^2=1$, which intersects every complex line that passes through the origin once,  except the line $\lambda^2=0$\footnote{Here upper index ''2'' refers to the second component of spinor $\lambda^\alpha$.}. The relation between the two representations is given by
\begin{equation}
\label{18feb3}
\begin{split}
\varphi(z,\bar z)=f(\lambda^1,1;\bar\lambda^{\dot 1},1), \qquad 
f(\lambda,\bar\lambda)=(\lambda^2)^N(\bar\lambda^{\dot 2})^{\bar N}\varphi\left(z,\bar z\right),\\
z\equiv \frac{\lambda^1}{\lambda^2}, \qquad \bar z\equiv \frac{\bar\lambda^{\dot 1}}{\bar\lambda^{\dot 2}}. \qquad\qquad\qquad\qquad\qquad
\end{split}
\end{equation}

\subsection{Deformed translations}

Having settled with the action of the Lorentz algebra, let us specify how the AdS deformed translations are realized. The $so(3,2)$ commutation relations require that these transform as vectors with respect to the Lorentz algebra. Accordingly, the result of the action of deformed translations on ${\cal V}^{(N,\bar N)}$ should belong to the tensor product
\begin{equation}
\label{1apr1}
{ P}{\cal V}^{(N,\bar N)}  \in {\cal V}^{(N,\bar N)} \otimes \Box_{so(3,1)},
\end{equation}
where $\Box_{so(3,1)}$ denotes the vector representation of $so(3,1)\sim sl(2,\mathbb{C})$. Tensor products of finite and infinite dimensional representations of the Lorentz algebra appeared in the context of weight-shifting operators \cite{Karateev:2017jgd,Costa:2018mcg}. For (\ref{1apr1}) the result is
\begin{equation}
\label{1apr2}
 {\cal V}^{(N,\bar N)} \otimes \Box_{so(3,1)} =  {\cal V}^{(N+1,\bar N+1)} \oplus  {\cal V}^{(N-1,\bar N-1)} \oplus  {\cal V}^{(N+1,\bar N-1)} \oplus  {\cal V}^{(N-1,\bar N+1)}.
\end{equation}
Accordingly, we end up with the following ansatz for the deformed translations
\begin{equation}
\label{4may2}
 {P}_{\alpha\dot\alpha}=\lambda_{\alpha}\bar\lambda_{\dot\alpha} A(\hat N, \hat {\bar N})+\frac{\partial}{\partial \lambda^{\alpha}}\frac{\partial}{\partial \bar\lambda^{\dot\alpha}}B(\hat N, \hat {\bar N})+
\lambda_\alpha \frac{\partial}{\partial \bar\lambda^{\dot\alpha}} C(\hat N, \hat {\bar N})+
\bar\lambda_{\dot\alpha} \frac{\partial}{\partial \lambda^{\alpha}} D(\hat N, \hat {\bar N}).
\end{equation}
These will be applied to ${\cal V}^{N,\bar N}$, so operators $\hat N$ and $\hat {\bar N}$ can be replaced with the associated weights of ${\cal V}^{N,\bar N}$.  

Representation (\ref{4may2}) for the deformed translation acting on $ {\cal V}^{(N,\bar N)}$ appears to have the same form as in the polynomial case \cite{Khabarov:2019dvi}. Still, we have a minor modification to the standard analysis. Namely,  $sl(2,\mathbb{C})$ representations 
${\cal V}^{(N,\bar N)}$ and ${\cal V}^{(-N-2,-\bar N-2)}$ are equivalent, the equivalence being established with the intertwining kernel
\begin{equation}
\label{18feb7}
\varphi^{'(-N-2,-\bar N-2)}(z')={\cal O}\varphi(z')=c\frac{i}{2}\int (z'-z)^{-N-2}(\bar z'-\bar z)^{-\bar N -2}\varphi^{(N,\bar N)}(z)dzd\bar z,
\end{equation}
where $c$ is an arbitrary complex coefficient\footnote{This is only true for generic $N$ and $\bar N$, see \cite{Gelfand} for details.}. Accordingly, ${\cal O}$ can be combined with any term in (\ref{4may2}) to give new terms, that transform as required for $ {P}$. At the same time, it is worth noting that, by the very same reason,  ${\cal O}$ can often be avoided. For example, one can use the equivalence ${\cal V}^{(-N-3,\bar N-3)}\sim {\cal V}^{(N+1,\bar N+1)}$ to replace
\begin{equation}
\label{18feb8}
P: \quad {\cal O}\lambda_\alpha\bar\lambda_{\dot\alpha}{\cal V}^{(N,\bar N)} \to {\cal V}^{(-N-3,-\bar N-3)}
\end{equation}
with
\begin{equation}
\label{18feb9}
P: \quad \lambda_\alpha\bar\lambda_{\dot\alpha}{\cal V}^{(N,\bar N)} \to {\cal V}^{(N+1,\bar N+1)},
\end{equation}
which does not contain ${\cal O}$. Operators of this type will still appear in the following in the context of separation of positive and negative energy modules.

\subsection{Commutator of deformed translations}
\label{sec:solve}
To make sure that translations (\ref{4may2}) give rise to the AdS${}_4$ isometry algebra $so(3,2)$, it remains to impose
\begin{equation}
\label{4may4}
\begin{split}
[ {P}_{\alpha\dot\alpha}, {P}_{\beta \dot\beta}]=-\frac{i}{R^2}\varepsilon_{\alpha\beta}\bar{ J}_{\dot\alpha\dot\beta}-
\frac{i}{R^2}\varepsilon_{\dot\alpha\dot\beta}{J}_{\alpha\beta},
\end{split}
\end{equation}
where $R$ is the AdS radius. By collecting the coefficients in front of 
\begin{equation}
\label{4may6}
\lambda_\alpha\lambda_\beta, \qquad \bar\lambda_{\dot\alpha}\bar\lambda_{\dot\beta}, \qquad 
\frac{\partial}{\partial \bar\lambda^{\dot\alpha}} \frac{\partial}{\partial \bar\lambda^{\dot\beta}}, 
\qquad \frac{\partial}{\partial\lambda^\alpha}\frac{\partial}{\partial \lambda^\beta},
\end{equation}
and setting them to zero, we obtain
\begin{equation}
\label{4may5}
\begin{split}
(\bar N+2) C(N+1,\bar N+1)A(N,\bar N)=
\bar N A(N+1,\bar N-1)C(N,\bar N),\\
(N+2) D(N+1,\bar N+1)A(N,\bar N)=
N A(N-1,\bar N+1)D(N,\bar N),\\
(N+2) B(N+1,\bar N-1)C(N,\bar N)=
N C(N-1,\bar N-1)B(N,\bar N),\\
(\bar N+2) B(N-1,\bar N+1)D(N,\bar N)=
\bar N D(N-1,\bar N-1)B(N,\bar N).
\end{split}
\end{equation}
Next, collecting the coefficients in front of ${ J}$ and $\bar { J}$ and setting them equal on the two sides of (\ref{4may4}), we find
\begin{equation}
\label{4may7}
\begin{split}
& -\bar N A(N -1,\bar N-1)B(N,\bar N)+
(\bar N+2)B(N+1,\bar N+1)A(N,\bar N)\\
& \qquad +
(\bar N+2)C(N-1,\bar N+1)D(N,\bar N)-
\bar N D(N+1,\bar N-1)C(N,\bar N)=-\frac{2}{R^2},\\
&-N A(N -1,\bar N-1)B(N,\bar N)+
(N+2)B(N+1,\bar N+1)A(N,\bar N)\\
&\qquad -NC(N-1,\bar N+1)D(N,\bar N)+
(N+2)D(N+1,\bar N-1)C(N,\bar N)=-\frac{2}{R^2}.
\end{split}
\end{equation}
To construct $so(3,2)$ modules, it only remains to solve (\ref{4may5}), (\ref{4may7}).

Before proceeding to the solution, we note that in the polynomial $sl(2,\mathbb{C})$ case some equations from  (\ref{4may5}), (\ref{4may7}) can drop out. 
This happens because  for polynomials vanishing homogeneity degree in $\lambda$ means that the function is independent of $\lambda$ and, accordingly, 
polynomial subspace in ${\cal V}^{(0,\bar N)}$ is annihilated by  $\frac{\partial}{\partial \lambda^{\alpha}}\frac{\partial}{\partial \lambda^{\beta}}$. This is not true for ${\cal V}^{(0,\bar N)}$ itself. Therefore, in the non-polynomial case all equations  (\ref{4may5}), (\ref{4may7}) should be kept. As we will see below, this will result in substantial modifications to the standard analysis.

\subsection{Solving consistency conditions}

A first obvious comment is that deformed translations relate ${\cal V}^{N,\bar N}$ with homogeneity degrees of the form
\begin{equation}
\label{4may8}
(N,\bar N)=(N^0+a, \bar N^0+b), \qquad a,b \in \mathbb{Z},  \qquad a+b =2\mathbb{Z}.
\end{equation}
In other words, irreducible representations of $so(3,2)$ may only involve weights that belong to a lattice of the form (\ref{4may8}). At the same time, not all weights of such a lattice have to belong to an irreducible representation. Indeed, as we will see below, some of the coefficients $A$, $B$, $C$ and $D$ can vanish, which means that different weight spaces ${\cal V}^{N,\bar N}$ are no longer related by the action of $ {P}$ and, hence, lattice (\ref{4may8}) may split into smaller lattices. 

By passing to new variables 
\begin{equation}
\label{4may9}
\begin{split}
F(N , \bar N)& \equiv B(N +1, \bar N+1)A(N , \bar N),\\
G(N , \bar N)& \equiv D(N +1, \bar N-1)C(N , \bar N),
\end{split}
\end{equation}
we can rewrite (\ref{4may5}) and (\ref{4may7}) as
\begin{equation}
\label{4may10}
(\bar N+2)(N+1)F(N ,\bar N)=\bar N(N +3)F(N +1,\bar N-1),
\end{equation}
\begin{equation}
\label{4may11}
(\bar N+2)(N +3)G(N +1,\bar N+1)=\bar N(N +1)G(N ,\bar N),
\end{equation}
\begin{equation}
\label{4may12}
\begin{split}
&-\bar N F(N -1,\bar N-1)+
(\bar N+2)F(N ,\bar N)\\
&\qquad \qquad\qquad +
(\bar N+2)G(N -1,\bar N+1)-
\bar NG(N ,\bar N)=-\frac{2}{R^2},
\end{split}
\end{equation}
\begin{equation}
\label{4may13}
\begin{split}
&-N  F(N -1,\bar N-1)+
(N +2)F(N ,\bar N)\\
&\qquad \qquad\qquad
-
N 
G(N -1,\bar N+1)+(N +2)G(N ,\bar N)=-\frac{2}{R^2}.
\end{split}
\end{equation}
 Once $F$ and $G$ are known, $A$, $B$, $C$ and $D$ can be found from (\ref{4may9}). This does not define the latter coefficients uniquely. However, it can be seen that this ambiguity is related to the possibility to rescale ${\cal V}^{N,\bar N}$ with different $N$ and $\bar N$ with independent coefficients. In other words, once this scaling freedom is fixed, (\ref{4may9}) allows to define $A$, $B$, $C$ and $D$ unambiguously. Hence, we can focus on solving (\ref{4may10})-(\ref{4may13}) for $F$ and $G$.

The process of solving (\ref{4may10})-(\ref{4may13}) is somewhat technical and not very instructive, so we just describe its key steps and give the end result. 
To start, one notices that once $F$ and $G$ are known for some general $(N,\bar N)=(N_0,\bar N_0)$, (\ref{4may10})-(\ref{4may13}) are sufficient to reconstruct $F$ and $G$ in the neighboring points of the $sl(2,\mathbb{C})$ weight lattice. Proceeding iteratively, one argues that in a similar manner one can reconstruct $F$ and $G$ for the whole lattice. One can design different iterative schemes and  these, in principle, may give different solutions.  If different iterative schemes end up giving different results, this implies that the initial data $F(N_0,\bar N_0)$, $G(N_0,\bar N_0)$ is too general and should be constrained to ensure consistency. 
Altogether, this means that once the initial data $F(N_0,\bar N_0)$, $G(N_0,\bar N_0)$ is fixed, there should be at most one solutions to  (\ref{4may10})-(\ref{4may13}). 

We followed a certain iterative scheme, which we do not specify here, an ended up with a solution
\begin{equation}
\label{5may3}
\begin{split}
\tilde F(N,\bar N)&=\tilde F(N_0,\bar N_0)\frac{x^2-y_0^2}{x_0^2-y_0^2}+\tilde G(N_0,\bar N_0)\frac{x^2-x^2_0}{y_0^2-x_0^2}+
(x^2-x_0^2)(x^2-y_0^2),\\
\tilde G(N,\bar N)&=\tilde G(N_0,\bar N_0)\frac{y^2-x_0^2}{y_0^2-x_0^2}+\tilde F(N_0,\bar N_0)\frac{y^2-y^2_0}{x_0^2-y_0^2}+
(y^2-y_0^2)(y^2-x_0^2),
\end{split}
\end{equation}
where 
\begin{equation}
\label{5may4}
\begin{split}
x&\equiv N+\bar N+3, \qquad \;\;\: y\equiv N-\bar N+1,\\
x_0&\equiv N_0+\bar N_0+3, \qquad y_0\equiv N_0-\bar N_0+1
\end{split}
\end{equation}
and $\tilde F$, $\tilde G$ are related to $F$, $G$ via 
\begin{equation}\label{4may17}
\begin{split}
F&=-\frac{1}{16 R^2}\frac{\tilde F}{(N+1)(N+2)(\bar N+1)(\bar N+2)}\\
G&=-\frac{1}{16 R^2}\frac{\tilde G}{(N+1)(N+2)\bar N(\bar N+1)}.
\end{split}
\end{equation}
One can check that (\ref{5may3}), indeed, solves (\ref{4may10})-(\ref{4may13}) for any initial data $F(N_0,\bar N_0)$, $G(N_0,\bar N_0)$. Combined with the previous discussion, this implies that (\ref{4may10})-(\ref{4may13}) has a unique solution for any initial data. Putting this differently,  we have found that (\ref{4may4}) has a two-parametric set of solutions and these can be given in the form (\ref{5may3}).

Now we would like to rewrite (\ref{5may3}) in a more symmetric form, that does not make any specific reference to the initial point of the iterative procedure. 
To this end we first note that, though, (\ref{5may3}) was derived for a single lattice of the form (\ref{4may8}), it has a natural analytic continuation to any $(N,\bar N)$. This turns $\tilde F$ and $\tilde G$ into polynomials of fourth degree in complex  variables -- $x$ and $y$ respectively. It is then convenient to choose for $(N_0,\bar N_0)$ one of the points, at which $\tilde F$ and $\tilde G$ vanish. This leads to
\begin{equation}
\label{5may3x1}
\begin{split}
\tilde F(N,\bar N)=
(x^2-x_0^2)(x^2-y_0^2),\\
\tilde G(N,\bar N)=
(y^2-y_0^2)(y^2-x_0^2).
\end{split}
\end{equation}
Here $x_0$ and $y_0$ may be regarded as two independent complex parameters of the solution. Clearly, these are completely interchangeable $(x_0,y_0)\sim(y_0,x_0)$. Moreover, since $x_0$ and $y_0$  enter (\ref{5may3x1}) only squared, without loss of generality we can assume that they both have non-negative real part.

\subsection{Symmetries of consistency conditions}
\label{sec:sym}

As a side remark, we note that consistency conditions (\ref{4may5}), (\ref{4may7}) have a simple symmetry relating its solutions. More precisely, due to the fact that
\begin{equation}
\label{26mar1}
\left[\frac{\partial}{\partial \lambda^\alpha},\lambda_\beta \right]=\varepsilon_{\beta\alpha}=
\left[\lambda_\alpha,\frac{\partial}{\partial \lambda^\beta}\right],
\end{equation}
replacement 
\begin{equation}
\label{26mar2}
\lambda_\alpha \leftrightarrow \frac{\partial}{\partial \lambda^{\alpha}}
\end{equation}
preserves commutation relations. This replacement leaves ${J}$ and $\bar{ J}$ intact and 
\begin{equation}
\label{26mar3}
N \to -N-2, \qquad \bar N \to \bar N.
\end{equation}
By considering the action of (\ref{26mar2}) on deformed translations, we find that $ {P}$ gets mapped to $ {P}'$ with the coefficient functions being
\begin{equation}
\label{26mar4}
\begin{split}
A'(-N-2,\bar N)=D(N,\bar N), \qquad B'(-N-2,\bar N)=C(N,\bar N),\\
C'(-N-2,\bar N)=B(N,\bar N), \qquad D'(-N-2,\bar N)=A(N,\bar N).
\end{split}
\end{equation}
Since (\ref{26mar2}) preserves commutation relations, we conclude that $ {P}'$ also defines a consistent $so(3,2)$ module. By computing the associated $F'$ and $G'$, it is not hard to see that these are given by the same $x_0$ and $y_0$ as those, defining the original module. Analogously, one can consider a map that replaces $\bar \lambda$ with $\frac{\partial}{\partial\bar\lambda}$. This observation will be used below to decrease the number of cases to analyse.

Besides that, symmetry (\ref{26mar2}) can point towards the existence of an intertwining kernel that relates equivalent representations. More precisely, for functions on $\mathbb{C}^2/ \{ 0\}$ one can suggest
\begin{equation}
\label{26mar5}
f'(\lambda',\bar\lambda')=\int d^2\lambda e^{\lambda_\beta \lambda^{'\beta}}f(\lambda,\bar\lambda)
\end{equation}
which obeys
\begin{equation}
\label{26mar6}
\begin{split}
\lambda'_{\alpha}f'(\lambda',\bar\lambda')=\int d^2\lambda e^{\lambda_\beta \lambda^{'\beta}}\frac{\partial}{\partial \lambda^{\alpha}}f(\lambda,\bar\lambda),\\
\frac{\partial}{\partial\lambda^{'\alpha}}f'(\lambda',\bar\lambda')=\int d^2\lambda e^{\lambda_\beta \lambda^{'\beta}}{ \lambda_{\alpha}}f(\lambda,\bar\lambda).
\end{split}
\end{equation}
Of course, one should require that (\ref{26mar5}) converges or can be consistently regularised. 

In the case when $so(3,2)$ module involves a finite set of $sl(2,\mathbb{C})$ weights,  the intertwining kernel (\ref{26mar5}) contains redundant integrations,  that arise due to constraints on the homogeneity degrees of $f$ in $\lambda$, $\bar\lambda$,
which, in turn, results in divergences. Instead of (\ref{26mar5}) for a state with weight $(N,\bar N)$ one can suggest an intertwiner
\begin{equation}
\label{26mar10}
 \varphi^{'(-N-2,\bar N)}(z',\bar z)= \Gamma(N+2) \int dz (z'-z)^{-N-2} \varphi^{(N,\bar N)}(z,\bar z),
\end{equation}
that maps it to the weight space $(-N-2,\bar N)$. A straightforward computation shows that it does formally act as required by (\ref{26mar2}). An important drawback of (\ref{26mar10}) is that for half-integer $N$ its integrands is double-valued. At the same time, by combining it with the complex conjugate operation, we end up with a single-valued integrand.
Clarification of the utility of these observations we leave for future research.

\section{Classification of short modules}
\label{sec:3}

Having the general $so(3,2)$ module in the manifestly $sl(2,\mathbb{C})$ covariant  form, we would like to study cases, in which it decomposes into submodules. We will be mostly interested in short modules, that is in situations in which modules develop submodules, supported on a finite domain of the $(N,\bar N)$ plane.
For this purpose representation (\ref{5may3x1}) is very convenient, as it indicates at what points $F$ and $G$ vanish and, hence, the truncation of the module is possible. To be more precise, we find that $F$ and $G$ vanish for
\begin{equation}
\label{24feb1}
\begin{split}
F(N,\bar N)&=0, \qquad \text{for} \qquad N+\bar N+3 = \pm x_0,\quad \cup \quad  N+\bar N+3 = \pm y_0\\
G(N,\bar N)&=0, \qquad \text{for} \qquad N-\bar N+1 = \pm x_0, \quad \cup \quad N-\bar N+1 = \pm y_0.
\end{split}
\end{equation}

Let us study more closely the factorisation patterns that (\ref{24feb1}) leads to.  Assume that the weight lattice has values $(N,\bar N)$ with $N+\bar N = a$ and $F$ vanishes at these points. Vanishing of $F(N,\bar N)=0$ entails that $A(N,\bar N)=0$ or/and  $B(N+1,\bar N+1)=0$, see (\ref{4may9}). For simplicity, we will assume that both $A$ and $B$ are vanishing. As it is not hard to see from the way  $ {P}$ is defined  (\ref{4may2}), this leads to the decoupling of the parent module into two submodules with 
\begin{equation}
\label{24feb2}
\begin{split}
N+\bar N &= a,\; a-2,\; a-4,\; \dots,\\
N+\bar N &= a+2,\; a+4,\; a+6,\; \dots.
\end{split}
\end{equation}
Similarly, if $G(N,\bar N)=0$ for $N-\bar N=b$ and these points belong to the lattice, then the module decomposes into submodules with weights
\begin{equation}
\label{24feb3}
\begin{split}
N-\bar N &= b,\; b-2,\; b-4,\; \dots,\\
N-\bar N &= b+2,\; b+4,\; b+6,\; \dots.
\end{split}
\end{equation}

 For a short module to arise, both $N+\bar N$ and $N -\bar N$ should be limited above and below to finite intervals. To limit $N+\bar N$ to a finite interval, in addition to $F=0$ for $N+\bar N=a$, we need to require $F=0$ for $N+\bar N =c$, 
\begin{equation}
\label{27febx1}
c=a+ 2i, \qquad i\in \mathbb{N}.
\end{equation}
 Then, the weight space consists of $i$ lines
\begin{equation}
\label{24feb4}
\begin{split}
N+\bar N= a+2,\; a+4,\; \dots, \; c.
\end{split}
\end{equation}
Analogously, if, in addition, $G=0$ for $N-\bar N =d$, 
\begin{equation}
\label{27febx2}
d=b+2j,\qquad  j\in \mathbb{N},
\end{equation}
 the weight space gets limited to $j$ lines in $N-\bar N$ variable
\begin{equation}
\label{24feb5}
\begin{split}
N-\bar N= b+2,\; a+4,\; \dots, \; d.
\end{split}
\end{equation}
As a result, the weight lattice reduces to a finite rectangle in the $(N, \bar N)$ plane.

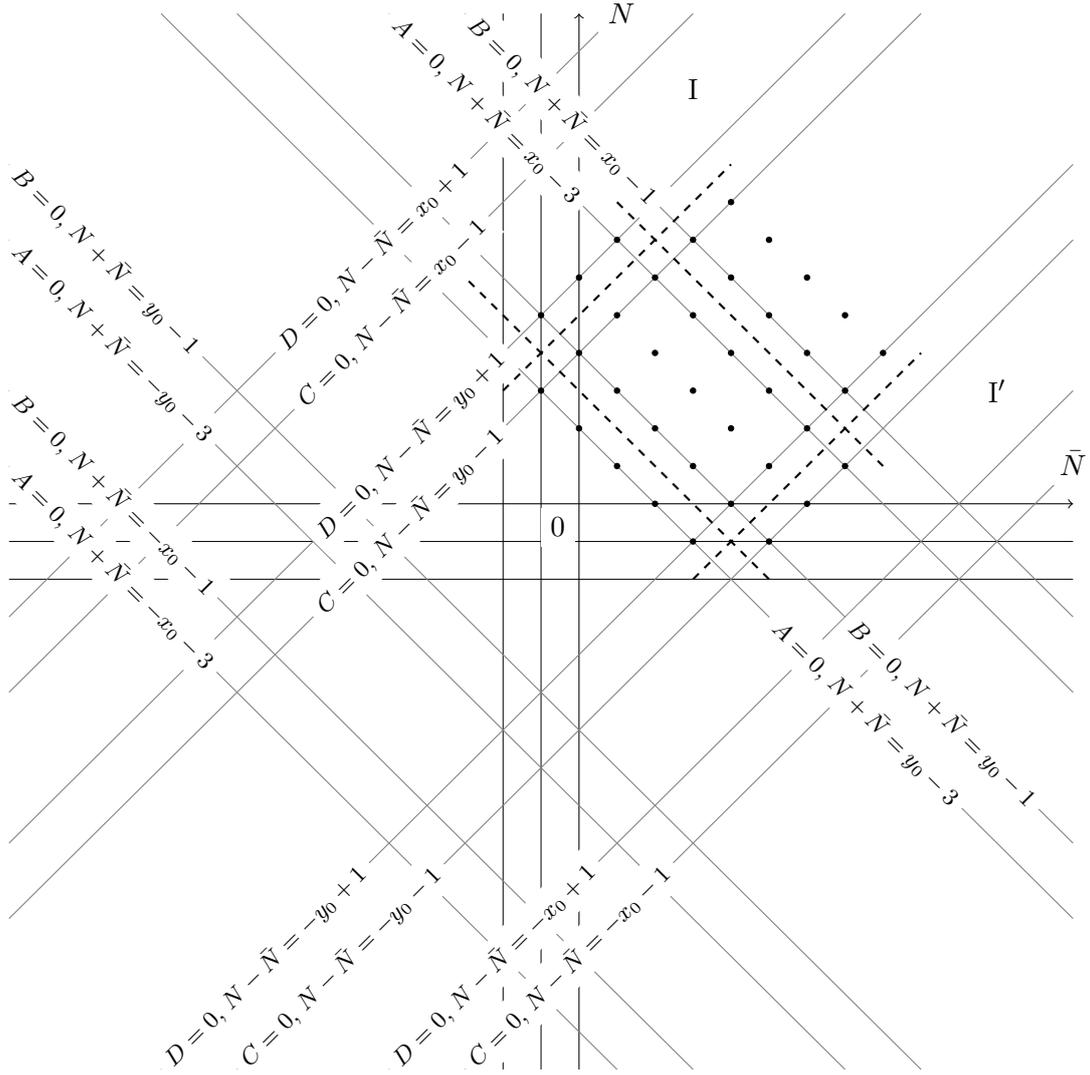
\begin{figure}
\centering
\begin{tikzpicture}[scale=0.5][>=stealth]
\draw [->] (-15,0) -- (13,0) ;
\draw [very thin] (-15,-1) --  (13,-1);
\draw [very thin] (-15,-2) -- (13,-2);
\draw  [->] (0,-15) -- (0,13);
\draw  [very thin] (-1,-15) -- (-1,13);
\draw  [very thin] (-2,-15) -- (-2,13);
\draw  [gray] (-5,13) --  (13,-5);
\draw [gray] (-3,13) -- (13,-3);
\draw [gray] (-15,1) --  (1,-15);
\draw [gray] (-15,3) -- (3,-15);
\draw [gray] (-15,-5) --  (3,13);
\draw [gray] (-15,-3) --  (1,13);
\draw [gray] (-3,-15) --  (13,1);
\draw [gray] (-5,-15) -- (13,3);
\draw [gray] (-11,13) --  (13,-11);
\draw(5,-3)  node[anchor=west,rotate around={-45:(0,0)},fill=white] {\footnotesize $A=0$, $N+\bar N=y_0-3$};
\draw [gray] (-9,13) -- (13,-9);
\draw(7,-3)  node[anchor=west,rotate around={-45:(0,0)},fill=white] {\footnotesize $B=0$, $N+\bar N=y_0-1$};
\draw [gray] (-15,7) --  (7,-15);
\draw(-15,7)  node[anchor=west,rotate around={-45:(0,0)},fill=white] {\footnotesize $A=0$, $N+\bar N=-y_0-3$};
\draw [gray] (-15,9) -- (9,-15);
\draw(-15,9)  node[anchor=west,rotate around={-45:(0,0)},fill=white] {\footnotesize $B=0$, $N+\bar N=y_0-1$};
\draw [gray] (-15,-11) -- (9,13);
\draw(-7,-3)  node[anchor=west,rotate around={45:(0,0)},fill=white] {\footnotesize $C=0$, $N-\bar N=y_0-1$};
\draw [gray] (-15,-9) -- (7,13);
\draw (-7,-1)  node[anchor=west,rotate around={45:(0,0)},fill=white] {\footnotesize $D=0$, $N-\bar N=y_0+1$};
\draw [gray] (-9,-15) --   (13,7);
\draw(-9,-15)  node[anchor=west,rotate around={45:(0,0)},fill=white] {\footnotesize $C=0$, $N-\bar N=-y_0-1$};
\draw [gray] (-11,-15) -- (13,9);
\draw(-11,-15)  node[anchor=west,rotate around={45:(0,0)},fill=white] {\footnotesize $D=0$, $N-\bar N=-y_0+1$};
\filldraw 
(4,2) circle (2pt)
(5,3) circle (2pt)
(3,3) circle (2pt)
(4,4) circle (2pt)
(2,4) circle (2pt)
(3,5) circle (2pt)
(1,5) circle (2pt)
(2,6) circle (2pt)
(6,4) circle (2pt)
(7,5) circle (2pt)
(5,5) circle (2pt)
(6,6) circle (2pt)
(4,6) circle (2pt)
(5,7) circle (2pt)
(3,7) circle (2pt)
(4,8) circle (2pt)
(0,4) circle (2pt)
(0,6) circle (2pt)
(1,3) circle (2pt)
(1,7) circle (2pt)
(-1,5) circle (2pt)
(-1,3) circle (2pt)
(0,2) circle (2pt)
(2,2) circle (2pt)
(3,1) circle (2pt)
(4,0) circle (2pt)
(5,1) circle (2pt)
(6,2) circle (2pt)
(7,3) circle (2pt)
(8,4) circle (2pt)
(1,1) circle (2pt)
(2,0) circle (2pt)
(3,-1) circle (2pt)
(5,-1) circle (2pt)
(6,0) circle (2pt)
(7,1) circle (2pt)
;
\draw(-5,13)  node[anchor=west,rotate around={-45:(0,0)},fill=white] {\footnotesize $A=0$, $N+\bar N=x_0-3$};
\draw(-3,13)  node[anchor=west,rotate around={-45:(0,0)},fill=white] {\footnotesize $B=0$, $N+\bar N=x_0-1$};
\draw(-15,1)  node[anchor=west,rotate around={-45:(0,0)},fill=white] {\footnotesize $A=0$, $N+\bar N=-x_0-3$};
\draw(-15,3)  node[anchor=west,rotate around={-45:(0,0)},fill=white] {\footnotesize $B=0$, $N+\bar N=-x_0-1$};
\draw(-7.5,2.5)  node[anchor=west,rotate around={45:(0,0)},fill=white] {\footnotesize $C=0$, $N-\bar N=x_0-1$};
\draw(-8,4)  node[anchor=west,rotate around={45:(0,0)},fill=white] {\footnotesize $D=0$, $N-\bar N=x_0+1$};
\draw(-3,-15)  node[anchor=west,rotate around={45:(0,0)},fill=white] {\footnotesize $C=0$, $N-\bar N=-x_0-1$};
\draw(-5,-15)  node[anchor=west,rotate around={45:(0,0)},fill=white] {\footnotesize $D=0$, $N-\bar N=-x_0+1$};
\draw(13,0.5) node[anchor=south,fill=white] {$\bar N$};
\draw(0.5,13)  node[anchor=west,fill=white] {$N$};
\draw(-0.1,-0.1) node[anchor=north east,fill=white] {$0$};
\draw[thick,dashed] (5,-2) -- (-3,6);
\draw[thick,dashed] (-2,3) -- (4,9);
\draw[thick,dashed] (1,8) -- (8,1);
\draw[thick,dashed] (3,-2) -- (9,4);
\draw(3,11) node {I};
\draw(11,3) node {I${}'$};
\end{tikzpicture}
\caption{This figure illustrates the way the decomposition of a parent $so(3,2)$ module into submodules may occur. Grey solid lines indicate points of the weight space for which coefficient functions $A$, $B$, $C$ and $D$ are equal to zero. Accordingly, the associated components of the deformed translations for these weights vanish. This gives various opportunities for $so(3,2)$ modules with a finite number of $sl(2,\mathbb{C})$ weights to occur. One such a module is shown above, its weights separated from the remaining weight lattice by two pairs of dashed lines. Two pairs of lines parallel to the coordinate axes indicate locations at which degeneracies of (\ref{4may5}), (\ref{4may7}) occur.}
\label{fig:1}
\end{figure}

Combining this general discussion with our results on zeroes of general solution (\ref{24feb1}), we find a multitude of opportunities for factorisations that result in short modules. Namely, we find
\begin{equation}
\label{27feb1}
\{ a, c\} = \{\pm x_0-3 ,\pm y_0-3\}, \qquad \{ b, d\} = \{\pm x_0-1 ,\pm y_0-1\},
\end{equation}
where it is meant that any of the variables on the left hand side of the equality can take any value on the right hand side, see Fig. \ref{fig:1}. 

Before proceeding, we need to take into account one subtlety, we were ignoring so far. Namely, for $N,\bar N = 0,-1,-2$, some coefficients in (\ref{4may5}), (\ref{4may10})-(\ref{4may13}) or  determinants relevant for the derivation of  the general solution vanish. Because of that, our passage from (\ref{4may5}) to (\ref{4may10}), (\ref{4may11}), that involved cancellations of these vanishing coefficients, may be invalid, as well as our general iterative scheme used to solve (\ref{4may10})-(\ref{4may13}) may fail. One manifestation of this problem is zeros in the denominators of (\ref{4may17}), that occur for certain values of $N$ and $\bar N$. 

To address this subtlety, one needs to carefully revisit our analysis at these special points. This boils down to a rather laborious case by case study, which we give in appendix \ref{App:A}. There we analyse all possibilities for $so(3,2)$ representations to have $sl(2,\mathbb{C})$ weight spaces ${\cal V}^{N,\bar N}$ with either $N$ or $\bar N$ taking values $0,-1,-2$. The conclusion of that analysis  is that these representations inevitably involve an infinite set of ${\cal V}^{N,\bar N}$ weights. Putting this differently, short representations of $so(3,2)$ cannot involve ${\cal V}^{N,\bar N}$ with special weights. 

This additional consideration significantly reduces the number of possibilities (\ref{27feb1}) for short modules to arise. Consider, for example, a case with
\begin{equation}
\label{27feb2}
a=y_0-3, \qquad d=y_0-1.
\end{equation}
In the corner of its weight space there is a point with
\begin{equation}
\label{27feb3}
N+\bar N = y_0 -1, \qquad N-\bar N = y_0-1, 
\end{equation}
which has $\bar N=0$. On Fig. \ref{fig:1} a highlighted rectangular domain has two corners at special points and, hence, it does not correspond to a consistent $so(3,2)$ module.

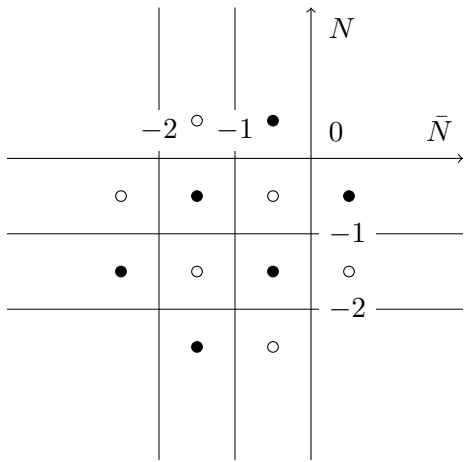
\begin{figure}
\centering
\begin{tikzpicture}[scale=1][>=stealth]
\draw [->] (-4,0) -- (2,0) ;
\draw [very thin] (-4,-1) -- (2,-1) ;
\draw [very thin] (-4,-2) -- (2,-2) ;
\draw  [->] (0,-4) -- (0,2);
\draw [very thin] (-1,-4) -- (-1,2) ;
\draw [very thin] (-2,-4) -- (-2,2) ;
\filldraw 
(-0.5,0.5) circle (2pt)
(0.5,-0.5) circle (2pt)
(-0.5,-1.5) circle (2pt)
(-1.5,-0.5) circle (2pt)
(-2.5,-1.5) circle (2pt)
(-1.5,-2.5) circle (2pt)
;
\draw (-0.5,-0.5) circle (2pt)
 (0.5,-1.5) circle (2pt)
  (-1.5,0.5) circle (2pt)
   (-2.5,-0.5) circle (2pt)
    (-1.5,-1.5) circle (2pt)
     (-0.5,-2.5) circle (2pt)
;
\draw(2,0.1) node[anchor=south east,fill=white] {$\bar N$};
\draw(0.1,2)  node[anchor=north west,fill=white] {$N$};
\draw(0.1,-1)  node[anchor=west,fill=white] {$-1$};
\draw(0.1,-2)  node[anchor=west,fill=white] {$-2$};
\draw(-1,0.1)  node[anchor=south,fill=white] {$-1$};
\draw(-2,0.1)  node[anchor=south,fill=white] {$-2$};
\draw(0.1,0.1)  node[anchor=south west,fill=white] {$0$};
\end{tikzpicture}
\caption{This figure shows rectangular lattices of $sl(2,\mathbb{C})$ weights for two short $so(3,2)$ modules: filled circles correspond to $i=3$, $j=2$, while empty circles correspond to $i=2$, $j=3$. Note that  both these solutions correspond to the same unordered pair $\{x_0,y_0 \} = \{2,3 \}$. Note also that thanks to the fact that $i+j$ is odd, values of $N$ and $\bar N$ in this module are half-integer and, as a result, special points are avoided.}
\label{fig:2}
\end{figure}

By studying various options, it is not hard to see that the only possibility for a short module to occur is 
\begin{equation}
\label{27feb4}
\begin{split}
a&= -x_0-3, \qquad c = x_0 -3,\\
b&=-y_0-1, \qquad d= y_0 -1,
\end{split}
\end{equation}
where we assumed that both $x_0$ and $y_0$ are positive.
Taking into account (\ref{27febx1}), (\ref{27febx2}), we find that $x_0 =i \in \mathbb{N}$ and $y_0=j\in \mathbb{N}$. Then, the weight space is spanned by points with
\begin{equation}
\label{27feb5}
\begin{split}
N+\bar N &= -i -1,-i+1,\dots, i-3,\\
N-\bar N &= -j+1,-j+3, \dots, j-1.
\end{split}
\end{equation}
It remains to note that $i$ and $j$ should be of opposite parity, 
\begin{equation}
\label{27feb6}
i+j \in 2\mathbb{N}+1,
\end{equation}
otherwise the module involves special points. Indeed, provided (\ref{27feb6}) is satisfied, both $N$ and $\bar N$ are half-integer for all weights in the module, and special weights $-2,-1,0$ are, thus, avoided. The case of $x_0=3$, $y_0=2$, which corresponds to two short modules with $\{i,j \}=\{2,3 \}$ and $\{i,j \}=\{3,2 \}$ is illustrated on Fig. \ref{fig:2}.

To summarise, we find that short $so(3,2)$ modules are classified by two natural numbers $i$ and $j$ of opposite parity. For each pair $(i,j)$ there is a single $so(3,2)$ module, which under restriction to $sl(2,\mathbb{C})$ reduces to a direct sum of modules ${\cal V}^{N,\bar N}$ with the weights belonging to rectangular lattice (\ref{27feb5}). In the next section we will study properties of these modules as well as identify them as the compact subalgebra $so(3)\oplus so(2)$ lowest-weight modules.

\section{Unitarity}
\label{sec:4}

In the present section we will find which of the short modules derived in the previous section are unitary. 

To start, we remind the reader that for an $so(3,2)$ module to be unitary, there should exist an $so(3,2)$ invariant positive definite sesquilinear form on its states
\begin{equation}
\label{15mar1}
\begin{split}
(\Phi,\Psi)&=(\Psi,\Phi)^*,\\
(\Phi,\xi_1 \Psi_1+\xi_2\Psi_2)&=\xi_1 (\Phi,\Psi_1)+\xi_2 (\Phi,\Psi_2),\\
(\eta_1\Phi_1+\eta_2\Phi_2,\Psi)&=\eta_1^* (\Phi_1,\Psi)+\eta_2^*(\Phi_2,\Psi),\\
(\Psi,\Psi)&>0,\qquad \text{for} \qquad \Psi \ne 0,\\
(\Phi,A\Psi) &= (A\Phi,\Psi), \qquad \text{for}\qquad A=\{{ J}, {P} \}.
\end{split}
\end{equation}

By construction, for short modules of $so(3,2)$, the representation space under restriction to $sl(2,\mathbb{C})$ splits into a finite set of irreducible representations. For (\ref{15mar1}) to be satisfied, each of these representations should be unitary as a representation of $sl(2,\mathbb{C})$.
This implies that we have the following two potential options for short unitary $so(3,2)$ modules. 
First option is 
\begin{equation}
\label{15mar2}
i=2, \qquad j=1.
\end{equation}
In this case the $so(3,2)$ module consists of two $sl(2,\mathbb{C})$ weights
\begin{equation}
\label{15mar3}
(N,\bar N) = \left(-\frac{3}{2},-\frac{3}{2}\right) \cup \left(-\frac{1}{2},-\frac{1}{2}\right),
\end{equation}
both belonging to the supplementary series. 
The second option is 
\begin{equation}
\label{15mar4}
i=1, \qquad j=2k, \qquad k\in \mathbb{N}.
\end{equation}
In this case the weight space is spanned by
\begin{equation}
\label{15mar5}
(N,\bar N) = \left(-k-\frac{1}{2},k-\frac{3}{2}\right) \cup \left(-k+\frac{1}{2},k-\frac{5}{2}\right)\cup  \dots \cup \left(k-\frac{3}{2},-k-\frac{1}{2}\right).
\end{equation}
All these representations belong to the principal series of $sl(2,\mathbb{C})$. For review on unitary representations of $sl(2,\mathbb{C})$, and respective invariant norms, see \cite{Gelfand}.

The $so(3,2)$-invariant norm can then be given as a linear combination of invariant norms for each $sl(2,\mathbb{C})$ weight space. To ensure that the former norm is positive definite, this linear combination should involve positive prefactors in front of each positive definite $sl(2,\mathbb{C})$ norm. Finally, for the resulting module to be unitary it remains to ensure that the latter requirement is compatible with self-adjointness of the deformed translations. Below we will use these considerations to check whether two cases of $so(3,2)$ modules are unitary or not as representations of $so(3,2)$.

\subsection{Case $i=2$, $j=1$}

To define the norm, we will use representation  (\ref{18feb3}). More specifically, we have
\begin{equation}
\label{15mar6}
f(\lambda,\bar\lambda)=(\lambda^2)^{-\frac{1}{2}}(\bar\lambda^{\dot 2})^{-\frac{1}{2}}\varphi^{\left(-\frac{1}{2},-\frac{1}{2}\right)}(z,\bar z)+
(\lambda^2)^{-\frac{3}{2}}(\bar\lambda^{\dot 2})^{-\frac{3}{2}}\varphi^{\left(-\frac{3}{2},-\frac{3}{2}\right)}(z,\bar z).
\end{equation}
In these terms, the $sl(2,\mathbb{C})$-invariant inner product reads\footnote{To avoid confusion, we emphasize that indices $1$ and $2$ for $\lambda$ and $\bar \lambda$ refer to different components of $\lambda^{\alpha}$ and $\bar\lambda^{\dot\alpha}$. In turn,  $1$ and $2$ in $z_1$ and $z_2$ is just an additional label that allows us to deal with two $z$'s.}
\begin{equation}
\label{15mar7}
\begin{split}
(\varphi,\psi)&=\alpha\left(-\frac{1}{2},-\frac{1}{2}\right) \int d^2z_1 d^2 z_2 |z_1-z_2|^{-3} \left( \varphi^{\left(-\frac{1}{2},-\frac{1}{2}\right)}(z_2)\right)^*
 \psi^{\left(-\frac{1}{2},-\frac{1}{2}\right)}(z_1)\\
 &+\alpha\left(-\frac{3}{2},-\frac{3}{2}\right) \int d^2z_1 d^2 z_2 |z_1-z_2|^{-1}\left( \varphi^{\left(-\frac{3}{2},-\frac{3}{2}\right)}(z_2)\right)^*
 \psi^{\left(-\frac{3}{2},-\frac{3}{2}\right)}(z_1).
\end{split}
\end{equation}
For this product to be positive definite, one has to require
\begin{equation}
\label{15mar8}
\alpha \left(-\frac{1}{2},-\frac{1}{2}\right) >0 ,\qquad \alpha \left(-\frac{3}{2},-\frac{3}{2}\right) <0. 
\end{equation}

It remains to impose self-adjointness of $ {P}_{\alpha\dot\alpha}$
\begin{equation}
\label{15mar9}
( {P}_{\alpha\dot\alpha} \varphi,\psi)=( \varphi,  {P}_{\alpha\dot\alpha}\psi).
\end{equation}
Let us consider, for definiteness, $ {P}_{1\dot 1}$. One has
\begin{equation}
\label{15mar10}
\begin{split}
 {P}_{1\dot 1} \varphi^{\left(-\frac{1}{2},-\frac{1}{2}\right)}& = A\left(-\frac{3}{2},-\frac{3}{2}\right) \varphi^{\left(-\frac{1}{2},-\frac{1}{2}\right)},\\
 {P}_{1\dot 1} \varphi^{\left(-\frac{3}{2},-\frac{3}{2}\right)}& = B\left(-\frac{1}{2},-\frac{1}{2}\right) \frac{\partial^2}{\partial z\partial \bar z}\varphi^{\left(-\frac{3}{2},-\frac{3}{2}\right)}.
\end{split}
\end{equation}
It is worth stressing that $ {P} \varphi^{\left(-\frac{1}{2},-\frac{1}{2}\right)}$ belongs to the space, in which $\varphi^{\left(-\frac{3}{2},-\frac{3}{2}\right)}$ takes values and similarly $ {P}$ maps $\varphi^{\left(-\frac{3}{2},-\frac{3}{2}\right)}$ to the weight space ${\cal V}^{\left(-\frac{1}{2},-\frac{1}{2}\right)}$.
Then 
\begin{equation}
\label{15mar11}
\begin{split}
&( {P}_{1\dot 1}\varphi ,\psi)=\alpha\left(-\frac{1}{2},-\frac{1}{2}\right) \int d^2z_1 d^2 z_2 |z_1-z_2|^{-3}\left({A \varphi}^{\left(-\frac{3}{2},-\frac{3}{2}\right)}(z_2)\right)^*
 \psi^{\left(-\frac{1}{2},-\frac{1}{2}\right)}(z_1)\\
&\quad +\alpha\left(-\frac{3}{2},-\frac{3}{2}\right) \int d^2z_1 d^2 z_2 |z_1-z_2|^{-1}\left(B\frac{\partial^2}{\partial z_2\partial \bar z_2} \varphi^{\left(-\frac{1}{2},-\frac{1}{2}\right)}(z_2)\right)^*
 \psi^{\left(-\frac{3}{2},-\frac{3}{2}\right)}(z_1).
\end{split}
\end{equation}
Analogously, one finds $( \varphi,  {P}_{\alpha\dot\alpha}\psi)$. Direct computation then gives that $ {P}_{\alpha\dot\alpha}$ is self-adjoint for
\begin{equation}
\label{15mar12}
\alpha\left(-\frac{3}{2},-\frac{3}{2}\right) B\left(-\frac{1}{2},-\frac{1}{2}\right)=4\alpha\left(-\frac{1}{2},-\frac{1}{2}\right) \left( A\left(-\frac{3}{2},-\frac{3}{2}\right) \right)^*.
\end{equation}
Lorentz invariance implies that the requirement that the remaining components of deformed translations are self-adjoint leads to the same condition.

It remains to note that solution (\ref{4may17}), (\ref{5may3x1}) implies
\begin{equation}
\label{15mar13}
F\left(-\frac{3}{2},-\frac{3}{2}\right) = B\left(-\frac{1}{2},-\frac{1}{2}\right) A\left(-\frac{3}{2},-\frac{3}{2}\right) =-\frac{4}{R^2}.
\end{equation}
Combining (\ref{15mar12}) and (\ref{15mar13}), we get
\begin{equation}
\label{15mar14}
B B^* = -\frac{16}{R^2} \frac{\alpha\left(-\frac{1}{2},-\frac{1}{2}\right)}{\alpha\left(-\frac{3}{2},-\frac{3}{2}\right)}.
\end{equation}
The left-hand-side of this equation is positive, which means that $\alpha\left(-\frac{1}{2},-\frac{1}{2}\right)$ and $\alpha\left(-\frac{3}{2},-\frac{3}{2}\right)$ should be of opposite signs, for the norm to be invariant with respect to deformed translations. By taking any positive $\alpha\left(-\frac{1}{2},-\frac{1}{2}\right)$ and then finding the associated $\alpha\left(-\frac{3}{2},-\frac{3}{2}\right)$ from (\ref{15mar14}), we obtain a positive definite inner product, see (\ref{15mar8}). Therefore, we found that the $so(3,2)$ invariant positive definite inner product for representation of $i=2$, $j=1$ does exist, therefore, this representation is unitary. As we will show below, it corresponds to the scalar singleton. 

\subsection{Case $i=1$, $j=2k$}

In this case the $sl(2,\mathbb{C})$ invariant inner product reads
\begin{equation}
\label{15mar15}
\begin{split}
(\varphi,\psi) &= \alpha\left(-k-\frac{1}{2},k-\frac{3}{2}\right) \frac{i}{2}\int  dz d\bar z \left(\varphi^{\left(-k-\frac{1}{2},k-\frac{3}{2}\right)}(z)\right)^*\psi^{\left(-k-\frac{1}{2},k-\frac{3}{2}\right)}(z)\\
&+\alpha\left(-k+\frac{1}{2},k-\frac{5}{2}\right) \frac{i}{2}\int  dz d\bar z \left(\varphi^{\left(-k+\frac{1}{2},k-\frac{5}{2}\right)}(z)\right)^*\psi^{\left(-k+\frac{1}{2},k-\frac{5}{2}\right)}(z)\\
&+\dots \\
&+\alpha\left(k-\frac{3}{2},-k-\frac{1}{2}\right) \frac{i}{2}\int  dz d\bar z \left(\varphi^{\left(k-\frac{3}{2},-k-\frac{1}{2}\right)}(z)\right)^*\psi^{\left(k-\frac{3}{2},-k-\frac{1}{2}\right)}(z).
\end{split}
\end{equation}
It is positive definite if
\begin{equation}
\label{15mar16}
\alpha (N,\bar N) >0 \qquad \forall N ,\bar N
\end{equation}
relevant for this module. 

Let us now consider constraints coming from self-adjointness of the deformed translations. We will focus on a pair of weights $(N,\bar N)$ and $(N+1,\bar N-1)$, both in the module, and components of the deformed momenta that map the associated weight spaces into each other. We have
\begin{equation}
\label{15mar17}
\begin{split}
 {P}_{1\dot 1} \varphi^{\left(N,\bar N\right)} &= -C\left(N,\bar N\right) \frac{\partial\varphi^{\left(N,\bar N\right)}}{\partial\bar z},\\
 {P}_{1\dot 1} \varphi^{\left(N+1,\bar N-1\right)}& = -D\left(N+1,\bar N-1\right) \frac{\partial\varphi^{\left(N+1,\bar N-1\right)}}{\partial z}.
\end{split}
\end{equation}
Here $ {P}_{1\dot 1} \varphi^{\left(N,\bar N\right)}$ is a state in the weight space ${\cal V}^{\left(N+1,\bar N-1\right)}$ and 
$ {P}_{1\dot 1} \varphi^{\left(N+1,\bar N-1 \right)}$ is a state in the weight space ${\cal V}^{\left(N,\bar N\right)}$.
Then, self-adjointness of $ {P}_{1\dot 1}$ leads to 
\begin{equation}
\label{15mar18}
\alpha(N+1,\bar N-1) \left(C\left(N,\bar N\right)  \right)^*=-\alpha(N,\bar N)D\left(N+1,\bar N-1\right).
\end{equation}
It can be rewritten as
\begin{equation}
\label{15mar20}
\alpha(N+1,\bar N-1) \left(C\left(N,\bar N\right)  \right)^*C\left(N,\bar N\right)=-\alpha(N,\bar N)G\left(N,\bar N\right).
\end{equation}
Taking into account (\ref{15mar16}), we find that positive definiteness of the norm (\ref{15mar15}) together with its invariance with respect to deformed translations requires
\begin{equation}
\label{15mar21}
G(N,\bar N)<0.
\end{equation}

General formula (\ref{4may17}), (\ref{5may3x1}) gives
\begin{equation}
\label{15mar22}
G(-\bar N-2,\bar N) = -\frac{1}{R^2} \frac{\left(\bar N-k+\frac{1}{2}\right)\left(\bar N+k+\frac{1}{2}\right)}{\bar N(\bar N+1)},
\end{equation}
where we used that all weights we are dealing with have $N+\bar N=-2$. Zeros of
\begin{equation}
\label{15mar23}
\left(\bar N-k+\frac{1}{2}\right)\left(\bar N+k+\frac{1}{2}\right) 
\end{equation}
 give locations at which module truncates in the weight space. Therefore,  for all $(N,\bar N)$ and $(N+1,\bar N-1)$ both being in the module, (\ref{15mar23}) has the same sign, which turns out to be ''minus''. So, (\ref{15mar22}) is negative when 
 \begin{equation}
 \label{15mar24}
 \bar N(\bar N+1)<0.
 \end{equation}
For  $\bar N$ half-integer, this is only true for $\bar N=-\frac{1}{2}$. The associated $G$ is responsible for components of the deformed  translations that map ${\cal V}^{\left(-\frac{3}{2},-\frac{1}{2}\right)}$ and ${\cal V}^{\left(-\frac{1}{2},-\frac{3}{2}\right)}$ to each other. Other weights should be absent from the module, because for them $G$ is positive and, hence, positivity of the norm is not compatible with its invariance with respect to the deformed translations. 

In summary, we find that only case $i=1$, $j=2$ corresponds to the unitary $so(3,2)$ module. It contains only two weight spaces -- ${\cal V}^{\left(-\frac{3}{2},-\frac{1}{2}\right)}$ and ${\cal V}^{\left(-\frac{1}{2},-\frac{3}{2}\right)}$ -- and the only non-vanishing coefficient function is
\begin{equation}
\label{15mar25}
G\left(-\frac{3}{2},-\frac{1}{2}\right)=-\frac{4}{R^2}.
\end{equation}
The positive definite and invariant norm is given by (\ref{15mar15}) with $k=1$, where $\alpha$ are positive and satisfy (\ref{15mar18}). As will be shown in the next section, this representation corresponds to the  Dirac spin-$\frac{1}{2}$ singleton. 

\section{Identification as lowest-weight modules and conformal fields}
\label{sec:5}

In this section we will identify short $so(3,2)$ modules we constructed previously as the lowest weight modules of the maximal compact subalgebra of $so(3,2)$, $so(3)\oplus so(2)$. In this approach one splits the algebra generators into compact generators --- energy
\begin{equation}
\label{16mar1}
E\equiv RP^0 = -RP_0
\end{equation}
and spatial rotations $L_{mn}$, $m,n = 1,2,3$ -- and non-compact generators --
\begin{equation}
\label{16mar2}
\begin{split}
J^{+m}\equiv J^{0m}+iJ^{4m}=J^{0m}+iRP^m,
\\
J^{-m}\equiv J^{0m}+iJ^{4m}=J^{0m}-iRP^m.
\end{split}
\end{equation}
The latter raise and lower energy according to
\begin{equation}
\label{16mar3}
[E,J^{+m}]=J^{+m} ,\qquad [E,J^{-m}]=-J^{-m}.
\end{equation}
One then constructs an $so(3,2)$ module from the lowest-weight vector space
\begin{equation}
\label{16mar4}
J^{-m}|E_0,s_0\rangle =0,
\end{equation}
 carrying a representation of $so(3)\oplus so(2)$ characterised by energy $E_0$ and spin $s_0$, by applying to it raising operators. To identify a representation constructed in this manner, it is sufficient to give $E_0$ and $s_0$ for its lowest energy state.

Our goal will be the opposite: we already have $so(3,2)$ modules constructed and we need to find $E_0$ and $s_0$ associated with them. To this end, we will solve (\ref{16mar4}) for the lowest weight space and find the associated eigenvalue of $E$. To find the $so(3)$ spin of the lowest weight space, one can find a state in $|E_0,s_0\rangle$ with lowest $J^{12}$. For this state eigenvalue of $J^{12}$ gives minus the spin of the $so(3)$ representation carried by $|E_0,s_0\rangle$. To find the lowest $J^{12}$ vector, we will utilize the $J^{12}$-lowering operator
\begin{equation}
{ J}^-\equiv (J^{23}+iJ^{13}), \qquad [J^{12},{ J}^-]=-{ J}^-,
\end{equation} 
which annihilates the lowest  $J^{12}$ state. Alternatively, spin $s_0$ can be derived from the dimension of the lowest energy space $|E_0,s_0\rangle$ as a vector space. Some useful formulae for carrying out this analysis can be found in appendix \ref{App:conv}.

\subsection{Case $i=2$, $j=1$: scalar singleton}
\label{sec:5.1}

We will now proceed along the lines sketched above in order to identify $E_0$ and $s_0$ for the $i=2$, $j=1$ $so(3,2)$ module.

To represent the state of the $i=1$, $j=2$ module we will use (\ref{15mar6}). In other words, each state is given by a pair of functions $\varphi^{\left(-\frac{1}{2},-\frac{1}{2}\right)}$ and $\varphi^{\left(-\frac{3}{2},-\frac{3}{2}\right)}$, each depending on two variables $z$ and $\bar z$. The action of ${ J}$ is given in (\ref{4may1}), while $ {P}$ is defined in (\ref{4may2}), where the only non-vanishing coefficient functions are $A$ and $B$, that satisfy (\ref{15mar13}). We will keep $A$ arbitrary\footnote{In principle, one can set it to any convenient value. Still, we keep this freedom explicit as it may be helpful in studying various flat space limits.}, while $B$ will be expressed in terms of $A$ by the latter formula.

 We then solve for the states, which are annihilated by $J^{-m}$ (\ref{16mar2}). To express $J^{-m}$ in terms of ${ J}$ and $ {P}$ we defined originally, we use the vector-spinor dictionary (\ref{16mar5}). The lowering operator $J^{-m}$ has three components. Each of them acting on a state in ${\cal V}^{\left(-\frac{1}{2},-\frac{1}{2}\right)}\oplus {\cal V}^{\left(-\frac{3}{2},-\frac{3}{2}\right)}$ produces contributions to ${\cal V}^{\left(-\frac{1}{2},-\frac{1}{2}\right)}$ and to ${\cal V}^{\left(-\frac{3}{2},-\frac{3}{2}\right)}$, both of which should be set to zero. Therefore, overall, we obtain six differential equations for $\varphi^{\left(-\frac{1}{2},-\frac{1}{2}\right)}$ and $\varphi^{\left(-\frac{3}{2},-\frac{3}{2}\right)}$. These can be solved and the result is
\begin{equation}
\label{16mar6}
\varphi^{\left(-\frac{1}{2},-\frac{1}{2}\right)}_{\rm lw}=\frac{c}{(2+2\bar z z)^{\frac{1}{2}}}, \qquad \varphi_{\rm lw}^{\left(-\frac{3}{2},-\frac{3}{2}\right)}=-\frac{2}{A R}\frac{c}{(2+2\bar z z)^{\frac{3}{2}}},
\end{equation}
where $c$ is an arbitrary constant. This result can be rewritten as 
\begin{equation}
\label{16mar7}
f_{\rm lw}=\frac{c}{(2 \lambda^1\bar\lambda^{\dot 1}+2 \lambda^2\bar\lambda^{\dot 2})^{\frac{1}{2}}}-\frac{2}{A R}\frac{c}{(2 \lambda^1\bar\lambda^{\dot 1}+2 \lambda^2\bar\lambda^{\dot 2})^{\frac{3}{2}}}
\end{equation}
in terms of homogeneous variables. Next, by a direct computation, we  find that this state is an eigenstate of energy with eigenvalue $\frac{1}{2}$
\begin{equation}
\label{16mar8}
E f_{\rm lw}=\frac{1}{2}f_{\rm lw}.
\end{equation}
Clearly, as the lowest energy space in a given case is a single vector, it may only have spin $0$. Some details of the computation that we sketched here can be found in Appendix \ref{App:B}.

Analogously, one shows that the $i=2$, $j=1$ module has a unique highest weight state 
\begin{equation}
\label{16mar17}
f_{\rm hw}=\frac{c}{(2 \lambda^1\bar\lambda^{\dot 1}+2 \lambda^2\bar\lambda^{\dot 2})^{\frac{1}{2}}}+\frac{2}{A R}\frac{c}{(2 \lambda^1\bar\lambda^{\dot 1}+2 \lambda^2\bar\lambda^{\dot 2})^{\frac{3}{2}}}
\end{equation}
with energy $-\frac{1}{2}$. 
In other words, we can identify the $i=2$, $j=1$ module as the direct sum of positive and negative energy modules of the scalar singleton. 
 
 Let us note that 
 \begin{equation}
 \label{16mar18}
  \lambda^1\bar\lambda^{\dot 1}+ \lambda^2\bar\lambda^{\dot 2}=(\sigma^0)_{\alpha\dot\alpha}\lambda^{\alpha}\bar\lambda^{\dot\alpha}=
  2 v^0_{\alpha\dot\alpha}\lambda^{\alpha}\bar\lambda^{\dot\alpha},
 \end{equation}
where $v^{\rm lw}_{\alpha\dot\alpha}$ is a spinor counterpart of $v^{{\rm lw}|i} = (\frac{1}{2},0,0,0)$, a four-momentum, that one can formally assign to the lowest weight state of the singleton module\footnote{By this we mean that in flat space the spatial components of momentum for the lowest energy state vanish. At the same time, its time component is fixed by (\ref{16mar8}).}.  In these terms (\ref{16mar7}) can be rewritten as
\begin{equation}
\label{16mar19}
f_{\rm lw}=\frac{c}{(4 
v^{\rm lw}_{\alpha\dot\alpha}\lambda^{\alpha}\bar\lambda^{\dot\alpha}
)^{\frac{1}{2}}}-\frac{2}{A R}\frac{c}{(4 
v^{\rm lw}_{\alpha\dot\alpha}\lambda^{\alpha}\bar\lambda^{\dot\alpha}
)^{\frac{3}{2}}}.
\end{equation}
By acting on this state with Lorentz transformations with real (pseudo)-angles, we find other states in the positive energy module
\begin{equation}
\label{16mar19x1}
f=\frac{c}{(4 
v_{\alpha\dot\alpha}\lambda^{\alpha}\bar\lambda^{\dot\alpha}
)^{\frac{1}{2}}}-\frac{2}{A R}\frac{c}{(4 
v_{\alpha\dot\alpha}\lambda^{\alpha}\bar\lambda^{\dot\alpha}
)^{\frac{3}{2}}},
\end{equation}
where $v$ now is any real future directed vector with $v^2 = -\frac{1}{4}$. One can also carry out a complex Lorentz transformation that maps $v^{\rm lw}$ to $-v^{\rm lw}$. Our expectation is that this should give a highest weight vector of the negative energy module. This expectation is consistent with the explicit computation (\ref{16mar17}).

\subsubsection{Splitting into positive and negative energies}

One may wonder whether positive and negative energy modes can be separated in this approach. As this separation is Lorentz invariant, it is natural to look for it in the form 
\begin{equation}
\label{16mar20}
\varphi^{\left(-\frac{3}{2},-\frac{3}{2}\right)}(z,\bar z)= {\cal O}\varphi^{\left(-\frac{1}{2},-\frac{1}{2}\right)} (z,\bar z)=d\frac{i}{2}\int |z-z_1|^{-3}\varphi^{(-\frac{1}{2},-\frac{1}{2})} (z_1,\bar z_1)dz_1 d\bar z_1.
\end{equation}
The operator ${\cal O}$ appearing on the right hand side is the standard intertwining operator that establishes  equivalence of two $sl(2,\mathbb{C})$ representations: 
${\cal V}^{\left(-\frac{1}{2},-\frac{1}{2}\right)}$ and  ${\cal V}^{\left(-\frac{3}{2},-\frac{3}{2}\right)}$. To fix the yet undetermined numerical coefficient $d$, we will require that (\ref{16mar20}) is also invariant with respect to deformed translations. In particular, we consider
\begin{equation}
\label{16mar21}
\begin{split}
\tilde\varphi^{\left(-\frac{1}{2},-\frac{1}{2}\right)}&\equiv P_{1\dot 1}\varphi^{\left(-\frac{3}{2},-\frac{3}{2}\right)} = A\varphi^{\left(-\frac{3}{2},-\frac{3}{2}\right)} ,\\
\tilde\varphi^{\left(-\frac{3}{2},-\frac{3}{2}\right)}&\equiv P_{1\dot 1}\varphi^{\left(-\frac{1}{2},-\frac{1}{2}\right)} =B\frac{\partial^2 \varphi^{\left(-\frac{1}{2},-\frac{1}{2}\right)}}{\partial z\partial \bar z}
\end{split}
\end{equation}
and require
\begin{equation}
\label{16mar22}
\tilde\varphi^{\left(-\frac{3}{2},-\frac{3}{2}\right)}(z,\bar z)= {\cal O}\tilde\varphi^{\left(-\frac{1}{2},-\frac{1}{2}\right)} (z,\bar z).
\end{equation}
Combined with (\ref{16mar20}) it entails
\begin{equation}
\label{16mar23}
\begin{split}
B\varphi^{\left(-\frac{1}{2},-\frac{1}{2}\right)}(z,\bar z)&=4A
d\frac{i}{2}\int dz_1 d\bar z_1 |z-z_1|^{-1}\\
&\qquad \qquad \qquad\cdot
 d\frac{i}{2}\int dz_2 d\bar z_2 |z_1-z_2|^{-3}\varphi^{\left(-\frac{1}{2},-\frac{1}{2}\right)}(z_2,\bar z_2).
\end{split}
\end{equation}
Employing the standard formula
\begin{equation}
\label{16mar24}
\left(\frac{i}{2}\right)^2\int dz_1 d\bar z_1 |z-z_1|^{-1} |z_1-z_2|^{-3}
=
 -4\pi^2 \delta (z-z_2)\delta(\bar z-\bar z_2),
\end{equation}
we find 
\begin{equation}
\label{16mar25}
d_{\pm}=\pm \frac{1}{2\pi AR}.
\end{equation}

In other words, we find two possibilities to isolate an invariant submodule inside module $i=2$, $j=1$ associated with two signs in (\ref{16mar25}). Clearly, these submodules correspond to positive and negative energy modules. To find out what sign correspond to what submodule, we can evaluate explicitly the integral (\ref{16mar20}) for $\varphi^{\left(-\frac{1}{2},-\frac{1}{2}\right)}_{\rm lw}$ and see whether it leads to (\ref{16mar7}) or (\ref{16mar17}). This is done in appendix \ref{App:C} and the result is that ''plus'' in (\ref{16mar25}) corresponds to the positive energy module. 

Finally, we note that we can realise  scalar singleton modules of definite energy on a single weight space ${\cal V}^{\left(-\frac{3}{2},-\frac{3}{2}\right)}$ by combining the standard action of $ {P}$ with ${\cal O}$
\begin{equation}
\label{2apr1}
\tilde  {P}f^{\left(-\frac{3}{2},-\frac{3}{2}\right)} \equiv  {\cal O} A\lambda_{\alpha}\bar\lambda_{\dot\alpha} f^{\left(-\frac{3}{2},-\frac{3}{2}\right)}.
\end{equation}
In a similar way one can realise scalar singleton modules of definite energy on a weight space ${\cal V}^{\left(-\frac{1}{2},-\frac{1}{2}\right)}$.

\subsection{Case $i=1$, $j=2$: spinor singleton}
\label{sec:5.2}

Here we give our results on the lowest weight analysis for the $i=1$, $j=2$ case. 

Similarly to the scalar singleton case, we use the following representation for states in the module
\begin{equation}
\label{17mar1}
f(\lambda,\bar\lambda)=(\lambda^2)^{-\frac{3}{2}}(\bar\lambda^{\dot 2})^{-\frac{1}{2}}\varphi^{\left(-\frac{3}{2},-\frac{1}{2}\right)}(z,\bar z)+
(\lambda^2)^{-\frac{1}{2}}(\bar\lambda^{\dot 2})^{-\frac{3}{2}}\varphi^{\left(-\frac{1}{2},-\frac{3}{2}\right)}(z,\bar z).
\end{equation}
We remind the reader that in this case only two coefficient functions are non vanishing and satisfy  (\ref{15mar25}),
\begin{equation}
\label{17mar2}
D\left(-\frac{1}{2},-\frac{3}{2}\right)C\left(-\frac{3}{2},-\frac{1}{2}\right)=-\frac{4}{R^2}.
\end{equation}

By solving the lowest weight conditions, we find the following two solutions
\begin{equation}
\label{17mar3}
f_{1|{\rm lw}}=\frac{c_1 \bar \lambda^{\dot 1}}{(2 \lambda^1\bar\lambda^{\dot 1}+2 \lambda^2\bar\lambda^{\dot 2})^{\frac{3}{2}}}+\frac{CR}{2}\frac{c_1 \lambda^2}{(2 \lambda^1\bar\lambda^{\dot 1}+2 \lambda^2\bar\lambda^{\dot 2})^{\frac{3}{2}}},
\end{equation}
\begin{equation}
\label{17mar4}
f_{2|{\rm lw}}=\frac{c_2 \bar \lambda^{\dot 2}}{(2 \lambda^1\bar\lambda^{\dot 1}+2 \lambda^2\bar\lambda^{\dot 2})^{\frac{3}{2}}}-\frac{CR}{2}\frac{c_2 \lambda^1}{(2 \lambda^1\bar\lambda^{\dot 1}+2 \lambda^2\bar\lambda^{\dot 2})^{\frac{3}{2}}}.
\end{equation}
These have the following properties
\begin{equation}
\label{17mar5}
\begin{split}
E f_{1|{\rm lw}} &= f_{1|{\rm lw}}, \qquad J^{12}f_{1|{\rm lw}} = -\frac{1}{2}f_{1|{\rm lw}},\\
E f_{2|{\rm lw}} &= f_{2|{\rm lw}}, \qquad J^{12}f_{{2|{\rm lw}}} = \frac{1}{2}f_{2|{\rm lw}}.
\end{split}
\end{equation}
Therefore, the $i=1$, $j=2$ case describes the spinor singleton module. 

As a historical remark, we note that the way the spinor singleton representation is constructed here is closely related to the one, that appeared at intermediate stages in the original paper by Dirac \cite{Dirac:1963ta}. Namely, deformed translations were realised on a single weight space ${\cal V}^{\left(-\frac{1}{2},-\frac{3}{2}\right)}$ as
\begin{equation}
\label{2apr2}
 {P}'_{\alpha\dot\alpha}f^{\left(-\frac{1}{2},-\frac{3}{2}\right)} = {\cal C}
 \bar\lambda_{\dot\alpha}\frac{\partial}{\partial \lambda^\alpha}f^{\left(-\frac{1}{2},-\frac{3}{2}\right)},
\end{equation}
where ${\cal C}$ is the operator of complex conjugation. Note that the presence of ${\cal C}$ implies that deformed translations (\ref{2apr2}) are not linear and hermitian operators, but rather anti-linear and anti-hermitian.

\subsection{Case $i=1$, $j=4$: spin-$\frac{3}{2}$ singleton}
\label{sec:5.3}

As the last explicit example we consider one of the non-unitary modules. In the $i=1$, $j=4$ case, the $sl(2,\mathbb{C})$ weight space has four points
\begin{equation}
\label{17mar6}
{\cal V}^{\left(-\frac{5}{2},\frac{1}{2}\right)}\oplus {\cal V}^{\left(-\frac{3}{2},-\frac{1}{2}\right)}\oplus 
{\cal V}^{\left(-\frac{1}{2},-\frac{3}{2}\right)}\oplus 
{\cal V}^{\left(\frac{1}{2},-\frac{5}{2}\right)}.
\end{equation}
Accordingly, we parametrise the states of the module as
\begin{equation}
\label{17mar7}
\begin{split}
f(\lambda,\bar\lambda)&=
(\lambda^2)^{-\frac{5}{2}}(\bar\lambda^{\dot 2})^{\frac{1}{2}}\varphi^{\left(-\frac{5}{2},\frac{1}{2}\right)}(z,\bar z)+
(\lambda^2)^{-\frac{3}{2}}(\bar\lambda^{\dot 2})^{-\frac{1}{2}}\varphi^{\left(-\frac{3}{2},-\frac{1}{2}\right)}(z,\bar z)\\
& \qquad\qquad+
(\lambda^2)^{-\frac{1}{2}}(\bar\lambda^{\dot 2})^{-\frac{3}{2}}\varphi^{\left(-\frac{1}{2},-\frac{3}{2}\right)}(z,\bar z)
+
(\lambda^2)^{\frac{5}{2}}(\bar\lambda^{\dot 2})^{-\frac{1}{2}}\varphi^{\left(\frac{1}{2},-\frac{5}{2}\right)}(z,\bar z).
\end{split}
\end{equation}
The only non-vanishing coefficient functions of deformed translations are constrained by
\begin{equation}
\label{17mar8}
\begin{split}
D\left(-\frac{3}{2},-\frac{1}{2}\right)C\left(-\frac{5}{2},\frac{1}{2}\right)&=G\left(-\frac{5}{2},\frac{1}{2}\right)=\frac{4}{R^2},\\
D\left(-\frac{1}{2},-\frac{3}{2}\right)C\left(-\frac{3}{2},-\frac{1}{2}\right)&=G\left(-\frac{3}{2},-\frac{1}{2}\right)=-\frac{16}{R^2},\\
D\left(\frac{1}{2},-\frac{5}{2}\right)C\left(-\frac{1}{2},-\frac{3}{2}\right)&=G\left(-\frac{1}{2},-\frac{3}{2}\right)=\frac{4}{R^2}.
\end{split}
\end{equation}
For brevity, we  denote
\begin{equation}
\label{17mar9}
C_1\equiv C\left(-\frac{5}{2},\frac{1}{2}\right), \qquad C_2\equiv C\left(-\frac{3}{2},-\frac{1}{2}\right),
\qquad C_3 \equiv C\left(-\frac{1}{2},-\frac{3}{2}\right)
\end{equation}
and $D$ will be eliminated in terms of the associated $C$ by means of (\ref{17mar8}).

Next, we proceed in the usual way. As a result, we find that the lowest weight space is spanned by four states
\begin{equation}
\label{19mar1}
\begin{split}
f_{1|{\rm lw}}&=\frac{ \left(\bar \lambda^{\dot 1}\right)^3}{(2 \lambda^1\bar\lambda^{\dot 1}+2 \lambda^2\bar\lambda^{\dot 2})^{\frac{5}{2}}}-\frac{C_1R}{2}\frac{ \left(\bar \lambda^{\dot 1}\right)^2\lambda^2}{(2 \lambda^1\bar\lambda^{\dot 1}+2 \lambda^2\bar\lambda^{\dot 2})^{\frac{5}{2}}}\\
&-\frac{C_1C_2R^2}{8}\frac{ \bar \lambda^{\dot 1}\left(\lambda^2\right)^2}{(2 \lambda^1\bar\lambda^{\dot 1}+2 \lambda^2\bar\lambda^{\dot 2})^{\frac{5}{2}}}
-\frac{C_1C_2C_3R^3}{16}\frac{\left(\lambda^2\right)^3}{(2 \lambda^1\bar\lambda^{\dot 1}+2 \lambda^2\bar\lambda^{\dot 2})^{\frac{5}{2}}},
\end{split}
\end{equation}
\begin{equation}
\label{19mar2}
\begin{split}
f_{2|{\rm lw}}&=\frac{ \left(\bar \lambda^{\dot 1}\right)^2\bar \lambda^{\dot 2}}{(2 \lambda^1\bar\lambda^{\dot 1}+2 \lambda^2\bar\lambda^{\dot 2})^{\frac{5}{2}}}
+\frac{C_1R}{6}\frac{ \left(\bar \lambda^{\dot 1}\right)^2\lambda^1-2\bar \lambda^{\dot 1}\bar \lambda^{\dot 2}\lambda^2}{(2 \lambda^1\bar\lambda^{\dot 1}+2 \lambda^2\bar\lambda^{\dot 2})^{\frac{5}{2}}}\\
&-\frac{C_1C_2R^2}{24}\frac{ \bar \lambda^{\dot 2}\left(\lambda^2\right)^2-2\bar\lambda^{\dot 1}\lambda^1\lambda^2 }{(2 \lambda^1\bar\lambda^{\dot 1}+2 \lambda^2\bar\lambda^{\dot 2})^{\frac{5}{2}}}
+\frac{C_1C_2C_3R^3}{16}\frac{\lambda^1\left(\lambda^2\right)^2}{(2 \lambda^1\bar\lambda^{\dot 1}+2 \lambda^2\bar\lambda^{\dot 2})^{\frac{5}{2}}},
\end{split}
\end{equation}
\begin{equation}
\label{19mar3}
\begin{split}
f_{3|{\rm lw}}&=\frac{ \left(\bar \lambda^{\dot 2}\right)^2\bar \lambda^{\dot 1}}{(2 \lambda^1\bar\lambda^{\dot 1}+2 \lambda^2\bar\lambda^{\dot 2})^{\frac{5}{2}}}
-\frac{C_1R}{6}\frac{ \left(\bar \lambda^{\dot 2}\right)^2\lambda^2-2\bar \lambda^{\dot 1}\bar \lambda^{\dot 2}\lambda^1}{(2 \lambda^1\bar\lambda^{\dot 1}+2 \lambda^2\bar\lambda^{\dot 2})^{\frac{5}{2}}}\\
&-\frac{C_1C_2R^2}{24}\frac{ \bar \lambda^{\dot 1}\left(\lambda^1\right)^2-2\bar\lambda^{\dot 2}\lambda^1\lambda^2 }{(2 \lambda^1\bar\lambda^{\dot 1}+2 \lambda^2\bar\lambda^{\dot 2})^{\frac{5}{2}}}
-\frac{C_1C_2C_3R^3}{16}\frac{\lambda^2\left(\lambda^1\right)^2}{(2 \lambda^1\bar\lambda^{\dot 1}+2 \lambda^2\bar\lambda^{\dot 2})^{\frac{5}{2}}},
\end{split}
\end{equation}
\begin{equation}
\label{19mar4}
\begin{split}
f_{4|{\rm lw}}&=\frac{ \left(\bar \lambda^{\dot 2}\right)^3}{(2 \lambda^1\bar\lambda^{\dot 1}+2 \lambda^2\bar\lambda^{\dot 2})^{\frac{5}{2}}}+\frac{C_1R}{2}\frac{ \left(\bar \lambda^{\dot 2}\right)^2\lambda^1}{(2 \lambda^1\bar\lambda^{\dot 1}+2 \lambda^2\bar\lambda^{\dot 2})^{\frac{5}{2}}}\\
&-\frac{C_1C_2R^2}{8}\frac{ \bar \lambda^{\dot 2}\left(\lambda^1\right)^2}{(2 \lambda^1\bar\lambda^{\dot 1}+2 \lambda^2\bar\lambda^{\dot 2})^{\frac{5}{2}}}
+\frac{C_1C_2C_3R^3}{16}\frac{\left(\lambda^1\right)^3}{(2 \lambda^1\bar\lambda^{\dot 1}+2 \lambda^2\bar\lambda^{\dot 2})^{\frac{5}{2}}}.
\end{split}
\end{equation}
These satisfy the following properties
\begin{equation}
\label{19mar5}
\begin{split}
E f_{1|{\rm lw}} &= f_{1|{\rm lw}}, \qquad J^{12}f_{1|{\rm lw}} = -\frac{3}{2}f_{1|{\rm lw}},\\
E f_{2|{\rm lw}} &= f_{2|{\rm lw}}, \qquad J^{12}f_{2|{\rm lw}} = -\frac{1}{2}f_{2|{\rm lw}},\\
E f_{3|{\rm lw}}& = f_{3|{\rm lw}}, \qquad J^{12}f_{3|{\rm lw}} = \frac{1}{2}f_{3|{\rm lw}},\\
E f_{4|{\rm lw}} &= f_{4|{\rm lw}}, \qquad J^{12}f_{4|{\rm lw}} = \frac{3}{2}f_{4|{\rm lw}}.
\end{split}
\end{equation}
Therefore, the $i=1$, $j=4$ module is a lowest weight representation with the lowest weight space of energy $1$ and spin $\frac{3}{2}$. As will be shown below, it corresponds to a spin-$\frac{3}{2}$ conformal field, with the action involving one derivative. 

\subsection{General case}

Instead of deriving  the lowest weight vectors and identifying their quantum numbers explicitly, as we did in the previous sections, we can find the lowest weight quantum numbers of the modules in question indirectly. 

Namely, we start by noting that the weight space of the Weyl module of a partially massless field of spin $s$ and depth $t$ -- which corresponds to areas I and I${}'$ on Fig \ref{fig:1} -- should have \cite{Skvortsov:2006at,Ponomarev:2010st,Khabarov:2019dvi}
\begin{equation}
\label{19mar6}
N-\bar N = \pm 2(t+1), \pm 2(t+2), \dots \pm 2s.
\end{equation}
This allows to establish a relation between parameters $x_0$, $y_0$ and $s$, $t$
\begin{equation}
\label{19mar7}
s=\frac{x_0-1}{2}, \qquad t=\frac{y_0-1}{2}.
\end{equation}
At the same time, for partially massless fields in AdS${}_4$
\begin{equation}
\label{19mar8}
m^2=-\frac{1}{R^2}(s-t-1)(s+t),
\end{equation}
where for $m^2$ a convention is used, in which the mass squared is counted from the point, at which the theory develops gauge invariance of a massless theory. This means that, the associated wave operator  reads
\begin{equation}
\label{19mar9}
\left(\Box - \frac{1}{R^2}[(s-2)(s+1)-s]-m^2 \right)h^{a(s)}=0,
\end{equation}
where $h$ is a symmetric traceless rank-$s$ $so(3,1)$ tensor field. In this way, we find how the wave operator is expressed in terms of parameters $x_0$ and $y_0$.

It is worth remarking that despite for establishing (\ref{19mar7}) we used partially massless fields, which correspond to very special values of $m^2$ and $t$ only  takes values $0,1,\dots, s-1$, this does not mean that this relation is only valid at these special points. Indeed, one can use other values of $t$ and then $t$ serves just as an alternative parameter to $m^2$. We used the case of partially massless fields for convenience: for these $m^2$ the module factorises, which is a phenomenon that is easy to identify in different approaches and, hence, to match the parameters involved. Alternatively, we could have proceeded more directly, for example, by computing the values of the Casimir operators in terms of $x_0$ and $y_0$ and then matching them with the known results in terms of $s_0$ and $m^2$.

Next, the standard formula allows us to relate the coefficient of the complete mass term in the wave equation with the energy of the lowest weight space in the module
\begin{equation}
\label{19mar10}
M^2=\frac{1}{R^2}\left(E_0(E_0-3)-s\right),
\end{equation}
where 
\begin{equation}
\label{19mar11}
M^2= \frac{1}{R^2}[(s-2)(s+1)-s]+m^2.
\end{equation}
Combining (\ref{19mar7}), (\ref{19mar8}), (\ref{19mar10}) and (\ref{19mar11}), we find
\begin{equation}
\label{19mar12}
E_0(E_0-3)=R^2 m^2+(s-2)(s+1)=\frac{y_0^2-9}{4}.
\end{equation}
Accordingly, we have two solutions for $E_0$ in terms of $y_0$
\begin{equation}
\label{19mar13}
E_0=\frac{3\pm y_0}{2}.
\end{equation}
In terms of lowest weight eigenvalues, the general solution (\ref{5may3x1}) can then be rewritten as 
\begin{equation}
\label{27mar1}
\begin{split}
\tilde F(N,\bar N)=
\big(x^2-(2s_0+1)^2)(x^2-(2E_0-3)^2\big),\\
\tilde G(N,\bar N)=
\big(y^2-(2s_0+1)^2)(y^2-(2E_0-3)^2\big).
\end{split}
\end{equation}

From the perspective of the previous discussion, the fact that the same $y_0$ corresponds to two different lowest weights $E_0$ follows from the existence of two possibilities of the weight lattice to decouple for a fixed $y_0$: due to $F=0$ on $N+\bar N=-y_0-3$ and on $N+\bar N=y_0-3$. For short modules for the lowest weight energy one has to take the smaller solution in (\ref{19mar13}).
In fact, the whole reason for shortening of these modules is that their parent modules admit submodules, which are also lowest weight ones with the energy corresponding to the larger root in (\ref{19mar13}), which are then quotiented out. 

In summary, we find that $i=y_0$, $j=x_0$ module has lowest weight space with
\begin{equation}
\label{19mar14}
E_0 = \frac{3-i}{2}, \qquad s_0=\frac{j-1}{2}.
\end{equation}
This is consistent with our explicit computations for the examples considered in sections \ref{sec:5.1}, \ref{sec:5.2} and \ref{sec:5.3}.

\subsection{Identification as conformal fields}

There is a vast literature on different types of conformal fields in different dimensions analyzed from different perspectives, see e.g. \cite{Drew:1980yk,Barut:1982nj,Deser:1983tm,Fradkin:1985am,Siegel:1988gd,Metsaev:1995jp,Iorio:1996ad,Erdmenger:1997wy,Segal:2002gd,Dobrev:2005bd,Marnelius:2009uw,Metsaev:2009ym,Vasiliev:2009ck,Bekaert:2013zya,Metsaev:2016oic,Kuzenko:2019ill}. We will compare our results with  \cite{Vasiliev:2009ck}, in which bosonic conformal fields of mixed symmetry type in arbitrary dimension were classified. 
 
 In $3d$ the conformal Lagrangians can be labelled by spin $s_0$ and $\kappa$, which is half the number of derivatives in the action.  The latter parameter is related to the energy of the lowest weight state $E_0$ or, equivalently, to the conformal dimension, via
\begin{equation}
\label{31mar1}
E_0 = \frac{3}{2}-\kappa.
\end{equation}
Both $s$ and $\kappa$ are integers and $s_0\ge 0$, $\kappa \ge 1$. The associated Lagrangians read
\begin{equation}
\label{31mar2}
\begin{split}
{\cal L}&=\frac{1}{2}\sum_{n=0}^N \frac{2^n (\kappa +1-n)_n}{n! (s-n)! \left(\kappa +s+\frac{1}{2}-n \right)_n}\partial^{b_1}\dots \partial^{b_n}\phi_{b_1\dots b_na_1\dots a_{s_0-n} }\\
&\qquad \qquad\qquad \qquad\qquad \qquad\qquad \qquad\qquad \qquad
\Box^{\kappa-n}
\partial^{c_1}\dots \partial^{c_n}\phi_{c_1\dots c_n}{}^{a_1\dots a_{s_0-n} },
\end{split}
\end{equation}
where 
\begin{equation}
\label{31mar3}
(p)_q\equiv \frac{\Gamma(p+q)}{\Gamma(p)}, \qquad N\equiv {\rm min}(s_0,\kappa)
\end{equation}
and $\phi$ are rank-$s_0$ traceless symmetric $so(2,1)$ tensors. It is trivial to see that our results -- sets $(E_0,s_0)$ -- match these if we take $i$ even and $j$ odd, which corresponds to the bosonic case. At the same time, in the fermionic case our classification  suggests that  for any half-integer spin and any odd number of derivatives there is an associated conformal Lagrangian. 

Field theoretic realisation (\ref{31mar2}) of conformal fields makes $iso(2,1)$ part of  $so(3,2)$ symmetry manifest. Instead, the construction developed in the previous sections makes $so(3,1)$ part of the conformal symmetry manifest. From this perspective, one can regard the latter approach as a deformation of the former one, in which 3d Minkowski space is replaced with dS${}_3$, the two space being conformally equivalent.

\section{Flat space limit}
\label{sec:6}

In this section we discuss the flat space limit, $so(3,2)\to iso(3,1)$ of the previously constructed solutions. In appendix \ref{App:flat} we also present  short solutions to the flat space consistency conditions, which do not have $so(3,2)$ counterparts. More comprehensive analysis of the $iso(3,1)$ modules in this formalism we leave for future research.

In the flat space limit general solution in the form (\ref{4may17}), (\ref{5may3x1}) vanishes due to the choice of parameters. To have a non-trivial flat space result we should first trade $y_0$ for $m^2$ using (\ref{19mar7}), (\ref{19mar8}) and then send $R\to \infty$. As a result we find
\begin{equation}
\label{27mar2}
\begin{split}
F(N,\bar N)&=\frac{m^2}{4} \frac{x^2-x_0^2}{(N+1)(N+2)(\bar N+1)(\bar N+2)},\\
G(N,\bar N)&=\frac{m^2}{4} \frac{y^2-x_0^2}{(N+1)(N+2)\bar N(\bar N+1)},
\end{split}
\end{equation}
where $m^2$ is the flat space mass squared, and relation between spin and $x_0$ (\ref{19mar7}) is still valid. From (\ref{27mar2}) it follows that for  $m^2\ne 0$ 
\begin{equation}
\label{27mar3}
\begin{split}
F(N,\bar N)=0, \qquad \text{for} \qquad N+\bar N+3 = \pm x_0,\\
G(N,\bar N)=0, \qquad \text{for} \qquad N-\bar N+1 = \pm x_0.
\end{split}
\end{equation}
Therefore, unlike for $so(3,2)$, in flat space, each coefficient function vanishes only on one pair of lines in the weight space. This, in principle,  allows to decouple a finite domain of the weight lattice and expect that it gives a short Poincare module. These modules, however, necessarily contain special points. It is not hard to verify that inhomogeneous constraints (\ref{4may5}) at these points are not satisfied, so the domain separated by (\ref{27mar3}) does not define a consistent Poincare module.

For $m^2=0$ coefficient functions $F$ and $G$ vanish. This does not yet mean that translations are realized trivially: for example, one can achieve $F=0$ by setting $B=0$ and keeping $A$ non-trivial. In fact, this is exactly how the massless fields are described: by setting in addition $C=D=0$, the weight space 
$\bar N-N=2s$  describes helicity $s$ massless field. The trivial solution to the Poincare consistency conditions $F=G=0$  leaves plenty of opportunities for modules to factorise and, hence, for short modules to occur. It seems, however, that these solutions typically lead to nilpotent momenta, which will be illustrated right below. A comprehensive analysis of possible factorisation patterns in the flat space case we leave for future research. 

As an illustration of the aforementioned phenomenon, we consider the flat space limit of the  scalar singleton. In the flat space limit (\ref{15mar13}) implies
\begin{equation}
\label{27mar4}
F\left(-\frac{3}{2},-\frac{3}{2}\right) = B\left(-\frac{1}{2},-\frac{1}{2}\right) A\left(-\frac{3}{2},-\frac{3}{2}\right) =0.
\end{equation}
To keep the action of translations non-trivial, we may proceed as in the case of massless fields and set
\begin{equation}
\label{27mar5}
B\left(-\frac{1}{2},-\frac{1}{2}\right)=0,\qquad A\left(-\frac{3}{2},-\frac{3}{2}\right) \ne 0.
\end{equation}
 This, however, implies that translations act in a nilpotent manner
 \begin{equation}
 \label{27mar6}
  {P}_{\alpha\dot\alpha} {P}_{\beta\dot\beta}\propto B\left(-\frac{1}{2},-\frac{1}{2}\right) A\left(-\frac{3}{2},-\frac{3}{2}\right)=0.
 \end{equation}
Considering that for massless fields the action of translations is not nilpotent, representation (\ref{27mar5}) does not seem to give a viable candidate for the flat space version of the Flato-Fronsdal theorem \cite{Flato:1978qz}.
Besides that, (\ref{27mar5}) leads to problems with unitarity. Namely, self-adjointness of $ {P}$, (\ref{15mar12}), combined with (\ref{27mar5}) leads to
\begin{equation}
\label{27mar7}
\alpha\left(-\frac{1}{2},-\frac{1}{2}\right) =0,
\end{equation}
which means that the subspace of states in ${\cal V}^{\left(-\frac{1}{2},-\frac{1}{2}\right)}$ should have vanishing norm. Factoring out these zero-norm states from the module, one ends up with the states in ${\cal V}^{\left(-\frac{3}{2},-\frac{3}{2}\right)}$ on which momenta act trivially\footnote{The fact that deformed translations for singletons trivialize in the flat-space limit is known for a long time, see e.g. \cite{Flato:1978qz,Flato:1980zk,Fronsdal:1986ui}.
It is also worth pointing out a recent suggestion to achieve a flat-space version of the Flato-Fronsdal theorem by replacing the tensor product of representations with its deformed version \cite{Iazeolla:2008ix}.}. 

To summarise, we find that in the flat limit of short representations, one can keep the action of translations non-trivial, however, these are still nilpotent. Moreover, the associated modules are non-unitary. These can be made unitary by factoring out zero-norm states. This in turn, leads to trivially realised translations. 
 Accordingly, naive flat space limit of short $so(3,2)$ representations does not produce  suitable candidates for a putative flat version of the Flato-Fronsdal theorem and, hence, cannot underly flat higher-spin holography.

\section{Conclusions}

In the present paper we classified short modules of $so(3,2)$ using the framework, that utilises $sl(2,\mathbb{C})$ spinors and, thus, makes the Lorentz invariance manifest. An interesting feature of this approach when applied to short modules is that it becomes necessary to deal with infinite-dimensional representations of $sl(2,\mathbb{C})$ realised by non-polynomial functions of $sl(2,\mathbb{C})$ spinors. We found that short modules can be labelled with two positive integer numbers of opposite parity, which can be connected to  spin and energy of the lowest weight state in a simple way (\ref{19mar14}).
In the bosonic case, for which the literature is available, our results agree with those obtained earlier. We also studied unitarity of these representation and in agreement with the literature found that, in agreement with the literature, only two of them -- free scalar and free spin-$\frac{1}{2}$ fermion with the standard kinetic terms -- are unitary.

Our results indicate relevance of infinite-dimensional $sl(2,\mathbb{C})$ modules for the description of  $so(3,2)$ invariant systems. It would be interesting to extend this analysis to other $so(3,2)$ representations, which, unlike short $so(3,2)$ modules, under restriction to $sl(2,\mathbb{C})$ decompose into an infinite number of non-polynomial $sl(2,\mathbb{C})$ representations. Though, being rather exotic, some of these representations may turn out to be unitary and have energy bounded from below. It would also be interesting to explore the role of the half-Fourier transform briefly mentioned in section \ref{sec:solve} in this construction.

Our main motivation was to explore potential avenues for higher-spin holography in flat space either as an independent construction or as the  flat limit of the AdS holography. As we argued in the introduction, spinor-helicity formalism is beneficial for higher-spin theories in flat space, so it seems natural to seek higher-spin holography within this approach. In this respect, in agreement with the literature, we find that the flat space limit of AdS holography cannot be smooth. More precisely, the action of deformed translations on the singleton module in the flat space limit becomes nilpotent and, as a result, singletons cannot feature any viable Flato-Fronsdal theorem analogous to that in AdS space. 

This issue seems to be inherent to flat space holography more generally.
Namely, as follows already from the classification by Wigner \cite{Wigner:1939cj}, there are no unitary irreducible representation of the Poincare algebra with the states labelled by two real variables and with translations represented non-trivially. These are supposed to be flat space counterparts of fundamental fields, out of which single-trace operators are constructed. The fact that such representations are missing, implies that flat space amplitudes, reinterpreted as boundary correlators, cannot have an underlying field-theoretic description.

This leaves us with two options to proceed. The first one is to ignore the absence of the underlying singleton construction in flat space and construct flat-space higher-spin amplitudes by taking the flat-space limit of their AdS counterparts constructed holographically.  As was already mentioned in the introduction, this approach allows to obtain higher-spin three-point amplitudes, which are consistent with those in the chiral higher spin theory. Going further, one can obtain higher-point amplitudes in flat space in the same way. Smoothness of the flat space limit for these amplitudes remains an open question.

Alternatively, one can try to carry out the analysis of short representations of $iso(3,1)$ not relying on contractions of short $so(3,2)$ representations.  In this regard, it is worth pointing out, that we suggested some short $iso(3,1)$ modules, that do not have $so(3,2)$ counterparts. The analysis of their properties as well as a more comprehensive analysis of short $iso(3,1)$ representations we leave for future research. It may also be interesting to consider various modifications of the problem. One of them that we find particularly interesting is to let momenta be complex, which, as known, allows to avoid triviality of the massless three-point scattering and may also be helpful to resolve degeneracies that we are encountering in our analysis.

\acknowledgments

We would like to thank E. Skvortsov for fruitful discussions on various subjects related to the paper. We are also grateful to V. Didenko, R. Metsaev and E. Skvortsov for comments on the draft. This work was supported by Russian Science Foundation Grant 18-12-00507.

\appendix

\section{Conventions}
\label{App:conv}

In this section we collect our conventions as well as some useful formulae on the conversion between vector and spinor indices.

We use the mostly plus signature $\eta = {\rm diag}(-,+,+,+)$. The Pauli matrices are
\begin{equation}
\label{5nov3}
\sigma^0 = 
\left(\begin{array}{cccc}
1 &&& 0\\
0 && &1
\end{array}\right), \quad \sigma^1 = 
\left(\begin{array}{cccc}
0 & && 1\\
1& && 0
\end{array}\right), 
\quad 
 \sigma^2 = 
\left(\begin{array}{ccc}
0 && -i\\
i&& 0
\end{array}\right), \quad
 \sigma^3 = 
\left(\begin{array}{ccc}
1 && 0\\
0&& -1
\end{array}\right).
\end{equation}
These can be used to convert a vector index to a pair of spinor ones 
\begin{equation}
\label{5nov4}
v_{\alpha\dot\alpha}\equiv v_a (\sigma^a)_{\alpha\dot\alpha}.
\end{equation}
To raise and lower spinor indices we use the following convention
 \begin{equation}
\label{5nov6}
\lambda^\alpha = \epsilon^{\alpha\beta} \lambda_\beta, \qquad \lambda_\beta=\epsilon_{\beta\gamma} \lambda^\gamma,
\end{equation}
where
\begin{equation}
\label{5nov7}
\epsilon^{\alpha\beta}=\epsilon^{\dot\alpha\dot\beta}=
\left(
\begin{array}{cccc}
0&&& 1\\
-1&&&0
\end{array}
\right) = -\epsilon_{\alpha\beta}=-\epsilon_{\dot\alpha\dot\beta}.
\end{equation}
The same rule is used to raise and lower indices of the Pauli matrices.
 Relation (\ref{5nov4}) can be inverted to give
 \begin{equation}
 \label{5nov8}
 v_a=-\frac{1}{2}(\sigma_a)^{\dot\alpha\alpha}v_{\alpha\dot\alpha}.
 \end{equation}
 To this end one needs to use
 \begin{equation}
 \label{5nov9}
 (\sigma^a)_{\alpha\dot\alpha}(\sigma_a)_{\beta\dot\beta}=-2\epsilon_{\alpha\beta}\epsilon_{\dot\alpha\dot\beta},
 \qquad
  (\sigma^a)^{\alpha\dot\alpha}(\sigma_a)^{\beta\dot\beta}=-2\epsilon^{\alpha\beta}\epsilon^{\dot\alpha\dot\beta}.
 \end{equation}
 Some other useful formulae in our conventions include
\begin{equation}
\label{5nov18}
\begin{split}
(\sigma_{a})_{\alpha\dot{\alpha}}({\sigma}^{b})^{\dot{\alpha}\alpha}=-2\delta^{b}_{a},\qquad
(\sigma_{a})_{\alpha\dot{\alpha}}({\sigma}^{a})^{\dot{\beta}\beta}=-2\delta^{\beta}_{\alpha}\delta^{\dot{\beta}}_{\dot{\alpha}},\\
({\sigma}^a)^{\dot\alpha\beta}(\sigma^b)_{\beta\dot\beta}+({\sigma}^b)^{\dot\alpha\beta}(\sigma^a)_{\beta\dot\beta}=-2\eta^{ab}\delta^{\dot{\alpha}}_{\dot{\beta}}.
\end{split}
\end{equation}

 For antisymmetric rank two tensor $J_{ab}=-J_{ba}$ one has 
 \begin{equation}
 \label{5nov10}
 J_{\alpha\dot\alpha,\beta\dot\beta}\equiv J^{ab}(\sigma_a)_{\alpha\dot\alpha}(\sigma_b)_{\beta\dot\beta}.
 \end{equation}
 One can then show that antisymmetry of $J^{ab}$ implies that $J_{\alpha\dot\alpha,\beta\dot\beta}$ is of the form
  \begin{equation}
 \label{5nov11}
 J_{\alpha\dot\alpha,\beta\dot\beta} = \epsilon_{\alpha\beta}\bar J_{\dot\alpha\dot\beta}+
 \epsilon_{\dot\alpha\dot\beta}J_{\alpha\beta},
 \end{equation}
 with $J_{\alpha\beta}$ and $\bar J_{\dot\alpha\dot\beta}$ symmetric.
  Here
 \begin{equation}
 \label{5nov12}
 \begin{split}
J_{\alpha\beta}=J_{\beta\alpha}\equiv \frac{1}{2}(\sigma_a)_\alpha{}^{\dot\gamma} (\sigma_b)_{\beta\dot\gamma}J^{ab},\\
\bar J_{\dot\alpha\dot\beta}=\bar J_{\dot\beta\dot\alpha} = \frac{1}{2}(\sigma_a)^\gamma{}_{\dot\alpha} (\sigma_b)_{\gamma\dot\beta}J^{ab}.
\end{split}
 \end{equation}
For real $J_{ab}$, $J_{\alpha\beta}$ and $\bar J_{\dot\alpha\dot\beta}$ are complex conjugate to each other. One can invert (\ref{5nov12}), which leads to
 \begin{equation}
\label{5nov13}
J_{ab} = \frac{1}{4} (\sigma_a)^{\dot\alpha\alpha}(\sigma_b)^{\dot\beta \beta}(\epsilon_{\alpha\beta}\bar J_{\dot\alpha\dot\beta}+
\epsilon_{\dot\alpha\dot\beta}J_{\alpha\beta}).
\end{equation}

By converting from vector to spinor conventions, we find the following relations between $so(3,2)$ generators in vector and spinor notations
\begin{equation}
\label{16mar5}
\begin{split}
P^0=-\frac{1}{2}(P_{1\dot 1}+P_{2\dot 2}),\\
P^1=\frac{1}{2}(P_{1\dot 2}+P_{2\dot 1}),\\
P^2=\frac{i}{2}(P_{1\dot 2}-P_{2\dot 1}),\\
P^3=\frac{1}{2}(P_{1\dot 1}-P_{2\dot 2}),\\
J^{01}=\frac{1}{4}(J_{11}+\bar J_{\dot 1\dot 1}-J_{22}-\bar J_{\dot 2\dot 2}),\\
J^{02}=\frac{i}4(J_{11}-\bar J_{\dot 1\dot 1}+J_{22}-\bar J_{\dot 2\dot 2}),\\
J^{03}=-\frac{1}{4}(J_{12}+J_{21}+\bar J_{\dot 1\dot 2}+\bar J_{\dot 2\dot 1}),\\
J^{12}=\frac{i}{4}(-J_{12}-J_{21}+\bar J_{\dot 1\dot 2}+\bar J_{\dot 2\dot 1}),\\
J^{13}=\frac{1}4(J_{11}+\bar J_{\dot 1\dot 1}+J_{22}+\bar J_{\dot 2\dot 2}),\\
J^{23}=\frac{i}{4}(J_{11}-\bar J_{\dot 1\dot 1}-J_{22}+\bar J_{\dot 2\dot 2}).
\end{split}
\end{equation}

\section{Analysis of special points}
\label{App:A}

Here we review the analysis of consistency conditions (\ref{4may5}), (\ref{4may7}) at special points. To start, we note that the system we are exploring has many symmetries. In particular, $N\to -N-2$, $\bar N\to -\bar N -2$, $N \leftrightarrow \bar N$, which can be applied independently and should be accompanied with the respective changes of the coefficient functions. These symmetries reduce the number of special examples that we need to check. The origin of this symmetry is explained in section \ref{sec:sym}.

\paragraph{Case 1} We start by considering an example, in which one coordinate is special, while the other one is not. For definiteness, we consider a point with $\bar N=0$ and $N=N_0$ with $N_0$ non-special and assume that it belongs to the module. Then (\ref{4may5}) entails
\begin{equation}
\label{27feb7}
\begin{split}
 C(N_0+1,1)A(N_0,0)=0,\\
 B(N_0-1,1)D(N_0,0)=
0.
\end{split}
\end{equation}
Here we have different options to set different multipliers to zero. 

\paragraph{Case 1.1} We start by considering an option
\begin{equation}
\label{27feb8}
\begin{split}
 C(N_0+1,1) =0, \qquad B(N_0-1,1)=0
 \end{split}
\end{equation}
with other coefficients non-zero. If this is the case, then points $(N_0-1,1)$ and $(N_0+1,1)$ both belong to the module, moreover,
\begin{equation}
\label{27feb9}
F(N_0-2,0)=0,\qquad G(N_0+1,1)=0
\end{equation}
as a consequence of (\ref{27feb8}). Points $(N_0-1,1)$ and $(N_0+1,1)$ are both non-special, so one can carry out the standard reconstruction procedure for $F$ and $G$ from these points to $N\ge 1$ as was done in section \ref{sec:solve}. This will lead to (\ref{4may17}), (\ref{5may3x1}). To make sure that (\ref{27feb9}) is satisfied, we need to fix $x_0$ and $y_0$ appropriately. It is not hard to see, that without loss of generality, this can be achieved by setting
\begin{equation}
\label{27feb10}
x_0=N_0+1.
\end{equation}
With this done, consistency conditions are satisfied for all non-special points in the module, $\bar N\ge 1$, and we still have one free parameter $y_0$ at our disposal.

Next, we consider other consistency conditions at $(N_0,0)$. Here again, we have different options. 

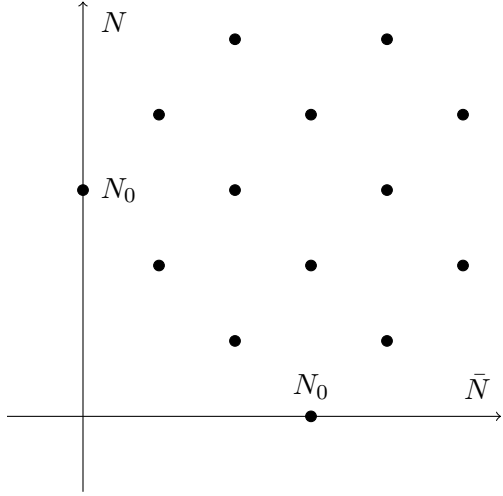
\begin{figure}
\centering
\begin{tikzpicture}[scale=1][>=stealth]
\draw [->] (-1,0) -- (5.5,0) ;
\draw  [->] (0,-1) -- (0,5.5);
\filldraw 
(0,3) circle (2pt)
(1,2) circle (2pt)
(2,1) circle (2pt)
(3,0) circle (2pt)
(1,4) circle (2pt)
(2,3) circle (2pt)
(3,2) circle (2pt)
(4,1) circle (2pt)
(2,5) circle (2pt)
(3,4) circle (2pt)
(4,3) circle (2pt)
(5,2) circle (2pt)
(4,5)  circle (2pt)
(5,4)  circle (2pt)
;
\draw(5.5,0.1) node[anchor=south east,fill=white] {$\bar N$};
\draw(0.1,5.5)  node[anchor=north west,fill=white] {$N$};
\draw(0.1,3)  node[anchor=west,fill=white] {$N_0$};
\draw(3,0.1)  node[anchor=south,fill=white] {$N_0$};
\end{tikzpicture}
\caption{This figure illustrates the weight space for Case 1.1.1. The non-vanishing coefficients entering the operators of deformed momenta are given by general formulae (\ref{4may17}), (\ref{5may3x1}) with $x_0=y_0=N_0+1$.}
\label{fig:3}
\end{figure}

\paragraph{Case 1.1.1} We will start from the one with
\begin{equation}
\label{27feb11}
B(N_0,0)=0,\qquad C(N_0,0)=0,
\end{equation}
or, in other words, the case of a module that does not extend to $\bar N<0$. Then, by straightforward computation, one sees that inhomogeneous equations 
(\ref{4may11}) at $(N_0,0)$ give
\begin{equation}
\label{27feb11x1}
\frac{1}{R^2}(y_0^2 - (N_0+1)^2)=0,
\end{equation}
 which, for finite $R$, without loss of generality, entails
\begin{equation}
\label{27feb12}
y_0=N_0+1.
\end{equation}
It is not hard to see that the module that involves $(N_0,0)$, and is truncated at lines associated with parameters (\ref{27feb10}), (\ref{27feb12}) is unbounded in the weight space in the $\bar N \to \infty$ direction. In other words, option (\ref{27feb11}) does not lead to a short module, see Fig. \ref{fig:3}.

\paragraph{Case 1.1.2} Let us consider the opposite situation
\begin{equation}
\label{27feb13}
B(N_0,0)\ne 0,\qquad C(N_0,0)\ne 0.
\end{equation}
Then both $(N_0-1,1)$ and $(N_0+1,-1)$ belong to the module.  Next, we look at  inhomogeneous consistency conditions (\ref{4may11}) at $(N_0-1,-1)$. These are two equations, which are only compatible if
\begin{equation}
\label{27feb14}
\frac{1}{R^2}(y_0^2 - (N_0-1)^2)=0.
\end{equation}
On the other hand, consistency of two inhomogeneous equations at $(N_0+1,-1)$ gives
\begin{equation}
\label{27feb15}
\frac{1}{R^2}(y_0^2 - (N_0+3)^2)=0.
\end{equation}
We obtained two inconsistent equations for $y_0$ -- (\ref{27feb14}) and (\ref{27feb15}) --, which means that option (\ref{27feb13}) does not result in a consistent $so(3,2)$ module. 

\paragraph{Case 1.1.3} Now, we consider the case, in which only one of the coefficients $B(N_0,0)$, $C(N_0,0)$ is non-zero. For definiteness, we consider
\begin{equation}
\label{4mar1}
B(N_0,0)\ne 0,\qquad C(N_0,0)=0,
\end{equation}
This implies $G(N_0,0)=0$ and inhomogeneous equations (\ref{4may11}) at $(N_0,0)$ fix
\begin{equation}
\label{4mar2}
F(N_0-1,-1)=\frac{y^2-(N_0+1)^2}{4N_0R^2}.
\end{equation}
After fixing $G(N_0-1,-1)=0$, which follows from (\ref{4may5}) at $(N_0,0)$, we consider inhomogeneous equations at $(N_0-1,-1)$, which turn out to be consistent only for 
\begin{equation}
\label{4mar3}
y_0=\pm (N_0-1).
\end{equation}
With (\ref{4mar3}) fixed, we find that two inhomogeneous equations at  $(N_0-1,-1)$ are equivalent and establish a linear relation between $F(N_0-2,-2)$ and $G(N_0-2,0)$. It turns out that both of them cannot be zero. The argument is similar to that we had for case (\ref{27feb13}). Namely, once $F(N_0-2,-2)\ne 0$ and $G(N_0-2,0)\ne 0$, homogeneous equations (\ref{4may5}) at $(N_0-1,-1)$ imply that $F(N_0-3,-1)\ne 0$, $G(N_0-3,-1)\ne 0$ and that point $(N_0-3,-1)$ is in the module. Considering inhomogeneous consistency conditions at this latter point, we find two equations, which are inconsistent with the values of parameters that we previously fixed (\ref{27feb10}), (\ref{4mar3}). Hence, both $F(N_0-2,-2)$ and $G(N_0-2,0)$ cannot be vanishing.

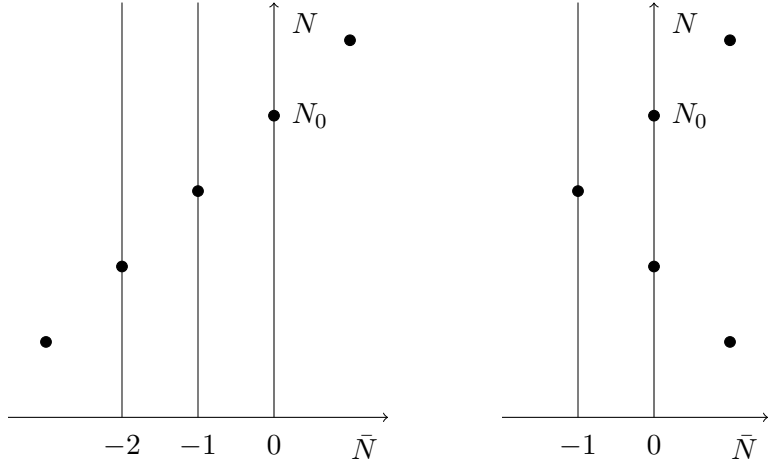
\begin{figure}
\centering
\begin{tikzpicture}[scale=1][>=stealth]
\draw  [->] (0,0) -- (0,5.5);
\draw  [very thin] (-1,0) -- (-1,5.5);
\draw  [very thin] (-2,0) -- (-2,5.5);
\draw [->] (-3.5,0) -- (1.5,0) ;
\filldraw 
(1,5) circle (2pt)
(0,4) circle (2pt)
(-1,3) circle (2pt)
(-2,2) circle (2pt)
(-3,1) circle (2pt)
;
\draw(1.5,-0.1) node[anchor=north east,fill=white] {$\bar N$};
\draw(0.1,5.5)  node[anchor=north west,fill=white] {$N$};
\draw(0.1,4)  node[anchor=west,fill=white] {$N_0$};
\draw(0,-0.1)  node[anchor=north,fill=white] {$0$};
\draw(-1,-0.1)  node[anchor=north,fill=white] {$-1$};
\draw(-2,-0.1)  node[anchor=north,fill=white] {$-2$};
\draw  [very thin] (4,0) -- (4,5.5);
\draw  [->] (5,0) -- (5,5.5);
\draw [->] (3,0) -- (6.5,0) ;
\filldraw 
(5,4) circle (2pt)
(6,5) circle (2pt)
(4,3) circle (2pt)
(5,2) circle (2pt)
(6,1) circle (2pt)
;
\draw(6.5,-0.1) node[anchor=north east,fill=white] {$\bar N$};
\draw(5.1,4)  node[anchor=west,fill=white] {$N_0$};
\draw(5.1,5.5)  node[anchor=north west,fill=white] {$N$};
\draw(5,-0.1)  node[anchor=north,fill=white] {$0$};
\draw(4,-0.1)  node[anchor=north,fill=white] {$-1$};
\end{tikzpicture}
\caption{Weight spaces for cases 1.1.3.1 (on the left) and 1.1.3.2 (on the right). All non-vanishing $F$ and $G$ are equal to $-1/R^2$.}
\label{fig:4}
\end{figure}

\paragraph{Case 1.1.3.1} Let  $G(N_0-2,0)$ be equal to zero, then
\begin{equation}
\label{4mar4}
F(N_0-2,-2) =-\frac{1}{R^2}
\end{equation}
and $(N_0-2,-2)$ is also in the module. This point is special and analogous to $(N_0,0)$ we started from. Applying similar arguments, we can use solution (\ref{4may17}), (\ref{5may3x1}) with the parameters set as in (\ref{27feb10}), (\ref{4mar3}) for $\bar N<-2$ and see that it is consistent with (\ref{4mar4}). Eventually, we find that the weight space of the $so(3,2)$ module in this case consists of ${\cal V}^{N,\bar N}$ with $(N,\bar N)=(N_0+i,i)$ and $i$ integer. Moreover, the only non-vanishing coefficients are $F(N_0+i,i)=-1/R^2$, see Fig. \ref{fig:4}.

\paragraph{Case 1.1.3.2} Analogously, we consider an option with $F(N_0-2,-2) =0$, which entails 
\begin{equation}
\label{4mar5}
G(N_0-2,0) =-\frac{1}{R^2}
\end{equation}
and $(N_0-2,0)$ is in the module. The latter point is again special and we can analyze the system in a way that we did with $(N_0,0)$. The conclusion of this analysis is that  the weight space  then consists of points $(N,\bar N)=(N_0+i-1,i-1)$  and  $(N,\bar N)=(N_0-j-2,j)$, where $i$, $j$ are non-negative integers. Moreover, $F(N_0+i-1,i-1)=-1/R^2$ and $G(N_0-j-2,j)=-1/R^2$ and all the remaining coefficients are vanishing. By looking at the weight space, one can see that this $so(3,2)$ module can be regarded as a previous one, with the points $\bar N<-1$ reflected with respect to the line $\bar N=-1$, see Fig. \ref{fig:4}.

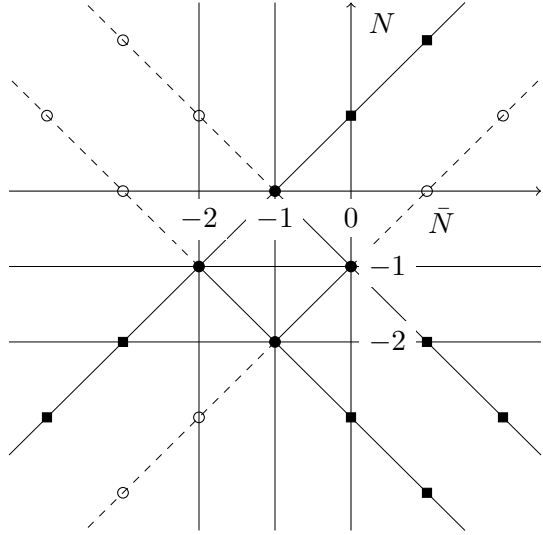
\begin{figure}
\centering
\begin{tikzpicture}[scale=1][>=stealth]
\draw  [->] (0,-4.5) -- (0,2.5);
\draw  [very thin] (-1,-4.5) -- (-1,2.5);
\draw  [very thin] (-2,-4.5) -- (-2,2.5);
\draw [->] (-4.5,0) -- (2.5,0) ;
\draw  [very thin] (-4.5,-1) -- (2.5,-1);
\draw  [very thin] (-4.5,-2) -- (2.5,-2);
\filldraw 
(-1,0) circle (2pt)
(0,-1) circle (2pt)
(-2,-1) circle (2pt)
(-1,-2) circle (2pt)
;
\filldraw
[xshift =1cm,yshift=2cm] (-0.06,-0.06) rectangle (0.06,0.06)
;
\filldraw
[xshift =0cm,yshift=1cm] (-0.06,-0.06) rectangle (0.06,0.06)
;
\filldraw
[xshift =-3cm,yshift=-2cm] (-0.06,-0.06) rectangle (0.06,0.06)
;
\filldraw
[xshift =-4cm,yshift=-3cm] (-0.06,-0.06) rectangle (0.06,0.06)
;
\filldraw
[xshift =1cm,yshift=-2cm] (-0.06,-0.06) rectangle (0.06,0.06)
;
\filldraw
[xshift =2cm,yshift=-3cm] (-0.06,-0.06) rectangle (0.06,0.06)
;
\filldraw
[xshift =0cm,yshift=-3cm] (-0.06,-0.06) rectangle (0.06,0.06)
;
\filldraw
[xshift =1cm,yshift=-4cm] (-0.06,-0.06) rectangle (0.06,0.06)
;
\draw 
(1,0) circle (2pt)
(2,1) circle (2pt)
(-2,-3) circle (2pt)
(-3,-4) circle (2pt)
(-2,1) circle (2pt)
(-3,2) circle (2pt)
(-3,0) circle (2pt)
(-4,1) circle (2pt)
;
\draw [very thin] (1.5,2.5) -- (-4.5,-3.5);
\draw [very thin] (-2,-1) -- (1.5,-4.5);
\draw [very thin] (-1,0) -- (2.5,-3.5);
\draw [very thin] (-1,-2) -- (0,-1);
\draw [very thin,dashed] (0,-1) -- (2.5,1.5);
\draw [very thin,dashed] (-1,-2) -- (-3.5,-4.5);
\draw [very thin,dashed] (-1,0) -- (-3.5,2.5);
\draw [very thin,dashed] (-2,-1) -- (-4.5,1.5);
\draw(1.5,-0.1) node[anchor=north east,fill=white] {$\bar N$};
\draw(0.1,2.5)  node[anchor=north west,fill=white] {$N$};
\draw(0,-0.1)  node[anchor=north,fill=white] {$0$};
\draw(-1,-0.1)  node[anchor=north,fill=white] {$-1$};
\draw(-2,-0.1)  node[anchor=north,fill=white] {$-2$};
\draw(0.1,-1)  node[anchor=west,fill=white] {$-1$};
\draw(0.1,-2)  node[anchor=west,fill=white] {$-2$};
\end{tikzpicture}
\caption{This picture represents a solution to consistency conditions (\ref{4may17}), (\ref{5may3x1}). The solution consists of weights, which are indicated by solid circles and solid squares. Non-vanishing $F$ and $G$ are represented by solid lines joining pairs of points. More precisely, non-vanishing $F(N,\bar N)$ is indicated by a line joining $(N,\bar N)$ and $(N+1,\bar N+1)$, while non-vanishing $G(N,\bar N)$ is denoted by a line that connects $(N,\bar N)$ and $(N+1,\bar N-1)$. The non-vanishing coefficients  are constrained by 
$F(-1,-2)+G(-1,0)=-1/R^2$, $F(-2,-1)=F(-1,-2)$ and $G(-2,-1)=G(-1,0)$. The remaining non-vanishing $F$ and $G$ are equal to $-1/R^2$.
There are also other solutions of a similar type. These can be obtained from the one above as follows. On the picture, there is a semi-infinite line of weights represented by solid squares, with the end-point at $(N,\bar N)=(-1,0)$. These weights can be replaced with those represented by empty circles and forming a line that also ends at $(-1,0)$. Alternatively, this can be described as the reflection of the weight space with respect to $N=-1$. After such a reflection a module is still consistent and the coefficient function on the new line ($F$ in this case) is also equal $-1/R^2$ at all points. There are four such independent reflections: they act on lines that end at $(-1,0)$, $(0,-1)$, $(-1,-2)$ and $(-2,-1)$.}
\label{fig:5}
\end{figure}

\begin{figure}
\centering
\begin{tikzpicture}[scale=1][>=stealth]
\draw  [->] (0,-1.5) -- (0,2.5);
\draw [->] (-1.5,0) -- (2.5,0) ;
\filldraw 
(0,0) circle (2pt)
(1,1) circle (2pt)
(2,2) circle (2pt)
;
\draw  [very thin] (0,0) -- (2.5,2.5);
\draw(2.5,-0.1) node[anchor=north east,fill=white] {$\bar N$};
\draw(0.1,2.5)  node[anchor=north west,fill=white] {$N$};
\draw(0.1,-0.1)  node[anchor=north west,fill=white] {$0$};
\draw(0.5,-5.5) node {a)};
\end{tikzpicture}
$\qquad\qquad$
\begin{tikzpicture}[scale=1][>=stealth]
\draw  [->] (0,-4.5) -- (0,2.5);
\draw  [very thin] (-1,-4.5) -- (-1,2.5);
\draw  [very thin] (-2,-4.5) -- (-2,2.5);
\draw [->] (-4.5,0) -- (2.5,0) ;
\draw  [very thin] (-4.5,-1) -- (2.5,-1);
\draw  [very thin] (-4.5,-2) -- (2.5,-2);
\filldraw 
(0,0) circle (2pt)
(1,1) circle (2pt)
(2,2) circle (2pt)
(-1,-1) circle (2pt)
(-2,-2) circle (2pt)
(-3,-3) circle (2pt)
(-4,-4) circle (2pt)
;
\draw  [very thin] (-4.5,-4.5) -- (2.5,2.5);
\draw(2.5,-0.1) node[anchor=north east,fill=white] {$\bar N$};
\draw(0.1,2.5)  node[anchor=north west,fill=white] {$N$};
\draw(0.1,-0.1)  node[anchor=north west,fill=white] {$0$};
\draw(-1,-0.1)  node[anchor=north,fill=white] {$-1$};
\draw(-2,-0.1)  node[anchor=north,fill=white] {$-2$};
\draw(0.1,-1)  node[anchor=west,fill=white] {$-1$};
\draw(0.1,-2)  node[anchor=west,fill=white] {$-2$};
\draw(-1,-5.5) node {b)};
\end{tikzpicture}\\
$\qquad$\\
$\qquad$\\
\begin{tikzpicture}[scale=1][>=stealth]
\draw  [->] (0,-4.5) -- (0,2.5);
\draw  [very thin] (-1,-4.5) -- (-1,2.5);
\draw  [very thin] (-2,-4.5) -- (-2,2.5);
\draw [->] (-4.5,0) -- (2.5,0) ;
\draw  [very thin] (-4.5,-1) -- (2.5,-1);
\draw  [very thin] (-4.5,-2) -- (2.5,-2);
\filldraw 
(0,0) circle (2pt)
(1,1) circle (2pt)
(2,2) circle (2pt)
(-1,-1) circle (2pt)
(-2,-2) circle (2pt)
(-3,-3) circle (2pt)
(-4,-4) circle (2pt)
(-2,0) circle (2pt)
(-3,1) circle (2pt)
(-4,2) circle (2pt)
(0,-2) circle (2pt)
(1,-3) circle (2pt)
(2,-4) circle (2pt)
;
\draw  [very thin] (-4.5,-4.5) -- (2.5,2.5);
\draw  [very thin] (-4.5,2.5) -- (2.5,-4.5);
\draw(2.5,-0.1) node[anchor=north east,fill=white] {$\bar N$};
\draw(0.1,2.5)  node[anchor=north west,fill=white] {$N$};
\draw(0.1,-0.1)  node[anchor=north west,fill=white] {$0$};
\draw(-1,-0.1)  node[anchor=north,fill=white] {$-1$};
\draw(-2,-0.1)  node[anchor=north,fill=white] {$-2$};
\draw(0.1,-1)  node[anchor=west,fill=white] {$-1$};
\draw(0.1,-2)  node[anchor=west,fill=white] {$-2$};
\draw(-1,-5.5) node {c)};
\end{tikzpicture}
\caption{Here we use the same conventions as for Fig \ref{fig:5} to indicate non-vanishing $F$ and $G$. On plot a) one has $F=-1/R^2$ for all $F$. On plot b) one has $F(-2,-2)+F(-1,-1)=-2/R^2$, while all the remaining $F$ are equal to $-1/R^2$. Finally, for plot c) one has $F(-2,2)+F(-1,-1)+G(-2,-0)+G(-1,-1)=-2/R^2$, while the remaining non-vanishing $F$ and $G$ are equal to $-1/R^2$.}
\label{fig:6}
\end{figure}

\paragraph{Other cases with non-special $N_0$} We have not yet studied all  options to set one or another coefficient to zero, but, as can be seen, they all reduce to what we have already considered or can be treated analogously. For example, if instead of (\ref{4mar1}) we consider 
\begin{equation}
\label{4mar5x1}
B(N_0,0) = 0,\qquad C(N_0,0)\ne 0,
\end{equation}
we get
\begin{equation}
\label{4mar6}
y_0=\pm (N_0+3).
\end{equation}
Then, the weight space consists of a line with $(N,\bar N)=(N_0-i+1,i-1)$ with $i$ non-negative integer, which can go either straight to $\bar N<-1$ or reflect against $\bar N = -1$. 

Another choice that we made was (\ref{27feb8}). Alternatively, one can solve (\ref{27feb7}) by setting 
\begin{equation}
\label{4mar7}
\begin{split}
 C(N_0+1,1) =0, \qquad D(N_0,0)=0.
 \end{split}
\end{equation}
In this case, the point $(N_0+1,1)$ belongs to the module. Then the first equation in (\ref{4mar7}) implies 
\begin{equation}
\label{4mar8}
G(N_0+1,1)=0
\end{equation}
while the second entails
\begin{equation}
\label{4mar8x1}
G(N_0-1,1)=0.
\end{equation}
These conditions fix
\begin{equation}
\label{4mar9}
x_0 =\pm (N_0+1), \qquad y_0 = \pm (N_0-1).
\end{equation}
This leads us to the setup of Case 1.1.3, which we already considered. Analogously, one considers 
\begin{equation}
\label{4mar10}
\begin{split}
 A(N_0,0) =0, \qquad B(N_0-1,1)=0
 \end{split}
\end{equation}
as a way to solve (\ref{27feb7}). The last remaining option to solve this equation is  
\begin{equation}
\label{4mar11}
\begin{split}
 A(N_0,0) =0, \qquad D(N_0,0)=0.
 \end{split}
\end{equation}
This entails 
\begin{equation}
\label{4mar12}
F(N_0,0)=0, \qquad G(N_0-1,1)=0,
\end{equation}
which are inconsistent with inhomogeneous equations at $(N_0,0)$ for $1/R^2 \ne 0$.

This completes the analysis of modules that involve  special points of the form $(N_0,0)$ with $N_0$ general. Special points $(N_0,-1)$ and $(N_0,-2)$ we considered along the way. Points for which $N$ takes special values, while $\bar N$ is general can be considered analogously. It remains to consider the situation, in which both $N$ and $\bar N$ take special values.

\paragraph{Case 2} Here we briefly discuss the case, for which both $N$ and $\bar N$ take special values. This analysis is very similar in spirit to what was done above: it amounts to the study of all consistency conditions at all points in the module. As this analysis is tedious and not very instructive, we just present its results. We focus only on solutions, which were not covered by the case with $\bar N_0$ special and $N_0$ any, that we considered before. These extra solutions to the consistency conditions are given on Fig \ref{fig:5} and Fig \ref{fig:6}. As mentioned at the beginning of this appendix, other solutions can be generated by transformations  $N\to -N-2$, $\bar N\to -\bar N -2$, $N \leftrightarrow \bar N$.

\section{Differential equations for lowest weight states}
\label{App:B}

In this Appendix we give some intermediate results of the derivation of the lowest weight states given in section \ref{sec:5.1}.

First we list contributions of $J^{-m}$ to each of the sectors, ${\cal V}^{(-\frac{1}{2},-\frac{1}{2})}$ and ${\cal V}^{(-\frac{3}{2},-\frac{3}{2})}$
\begin{equation}
\label{16mar9}
\begin{split}
&J^{-1}f\Big|_{{\cal V}^{(-\frac{3}{2},-\frac{3}{2})}}=\frac{i}{4AR} \big( 3 A R(z+\bar z)\varphi^{(-\frac{3}{2},-\frac{3}{2})}
+2AR(\bar z^2 -1)\bar\partial\varphi^{(-\frac{3}{2},-\frac{3}{2})}\\
&\quad-4 \bar\partial\varphi^{(-\frac{1}{2},-\frac{1}{2})}
+2AR(z^2-1)\partial \varphi^{(-\frac{3}{2},-\frac{3}{2})}-4\partial \varphi^{(-\frac{1}{2},-\frac{1}{2})}- 8(z+\bar z)\partial \bar\partial \varphi^{(-\frac{1}{2},-\frac{1}{2})}\big)
\end{split}
\end{equation}
\begin{equation}
\label{16mar10}
\begin{split}
&J^{-1}f\Big|_{{\cal V}^{(-\frac{1}{2},-\frac{1}{2})}}=\frac{i}{4} \big(2AR(z+\bar z) \varphi^{(-\frac{3}{2},-\frac{3}{2})}
+ (z+\bar z) \varphi^{(-\frac{1}{2},-\frac{1}{2})}\\
& \qquad\qquad\qquad\qquad\qquad+
2(\bar z^2 -1) \bar\partial\varphi^{(-\frac{1}{2},-\frac{1}{2})}
+2 (z^2 -1)  \partial\varphi^{(-\frac{1}{2},-\frac{1}{2})}
 \big)
\end{split}
\end{equation}
\begin{equation}
\label{16mar11}
\begin{split}
&J^{-2}f\Big|_{{\cal V}^{(-\frac{3}{2},-\frac{3}{2})}}=
-\frac{3}{4}(\bar z-z)\varphi^{(-\frac{3}{2},-\frac{3}{2})}
+\frac{1}{2AR}\big(
-AR(\bar z^2+1) \bar\partial \varphi^{(-\frac{3}{2},-\frac{3}{2})}
\\
&\quad - 2\bar\partial \varphi^{(-\frac{1}{2},-\frac{1}{2})}
+ AR(z^2+1) \partial \varphi^{(-\frac{3}{2},-\frac{3}{2})}
+2 \partial \varphi^{(-\frac{1}{2},-\frac{1}{2})}
+4(\bar z - z)\partial \bar\partial  \varphi^{(-\frac{1}{2},-\frac{1}{2})}
\big)
\end{split}
\end{equation}
\begin{equation}
\label{16mar12}
\begin{split}
&J^{-2}f\Big|_{{\cal V}^{(-\frac{1}{2},-\frac{1}{2})}}=
\frac{1}{4}\big(2AR (z-\bar z) \varphi^{(-\frac{3}{2},-\frac{3}{2})}
+(z-\bar z)\varphi^{(-\frac{1}{2},-\frac{1}{2})}\\
&\quad\qquad \qquad\qquad\qquad -
2(\bar z^2+1)\bar\partial \varphi^{(-\frac{1}{2},-\frac{1}{2})} +
2( z^2+1)\partial \varphi^{(-\frac{1}{2},-\frac{1}{2})}
\big)
\end{split}
\end{equation}
\begin{equation}
\label{16mar13}
\begin{split}
&J^{-3}f\Big|_{{\cal V}^{(-\frac{3}{2},-\frac{3}{2})}}=
-\frac{i}{2AR}\big( 
3 AR \varphi^{(-\frac{3}{2},-\frac{3}{2})}+\varphi^{(-\frac{1}{2},-\frac{1}{2})}
+2
AR\bar z\bar\partial \varphi^{(-\frac{3}{2},-\frac{3}{2})}\\
&\quad +2
\bar z\bar\partial \varphi^{(-\frac{1}{2},-\frac{1}{2})}
+2AR z \partial  \varphi^{(-\frac{3}{2},-\frac{3}{2})}
+ 2z \partial \varphi^{(-\frac{1}{2},-\frac{1}{2})}
+4(z\bar z-1)\partial\bar\partial 
\varphi^{(-\frac{1}{2},-\frac{1}{2})}
\big)
\end{split}
\end{equation}
\begin{equation}
\label{16mar14}
\begin{split}
&J^{-3}f\Big|_{{\cal V}^{(-\frac{1}{2},-\frac{1}{2})}}=
\frac{i}{2}\big(
AR (z\bar z-1) \varphi^{(-\frac{3}{2},-\frac{3}{2})}-
 \varphi^{(-\frac{1}{2},-\frac{1}{2})}\\
&\quad\qquad \qquad\qquad\qquad \qquad\qquad -2\bar z \bar\partial  \varphi^{(-\frac{1}{2},-\frac{1}{2})}-2z\partial  \varphi^{(-\frac{1}{2},-\frac{1}{2})}
\big).
\end{split}
\end{equation}
All these have to be set to zero. Thus, we obtain six equations. 

Equation that originates from (\ref{16mar10}) does not involve derivatives of $\varphi^{(-\frac{3}{2},-\frac{3}{2})}$ and can be easily solved for it. By plugging it into (\ref{16mar12}) and (\ref{16mar14}) we find two equations on $\varphi^{(-\frac{3}{2},-\frac{3}{2})}$ and its first derivatives. These two equations can be solved algebraically for the derivatives, which leads to
\begin{equation}
\label{16mar15}
\bar\partial \varphi^{(-\frac{1}{2},-\frac{1}{2})} = - \frac{z}{2 (1+z\bar z)}\varphi^{(-\frac{1}{2},-\frac{1}{2})},\qquad
\partial \varphi^{(-\frac{1}{2},-\frac{1}{2})} = - \frac{\bar z}{2 (1+z\bar z)}\varphi^{(-\frac{1}{2},-\frac{1}{2})}.
\end{equation}
These can be easily solved and lead to
\begin{equation}
\label{16mar16}
 \varphi^{(-\frac{1}{2},-\frac{1}{2})} =\frac{c}{(2+2z \bar z)^{\frac{1}{2}}}.
\end{equation}
Then, $\varphi^{(-\frac{3}{2},-\frac{3}{2})}$ can be reconstructed by setting the right hand side of (\ref{16mar10}) to zero. Then one can check that all the equations are satisfied.

\section{Lowest weight integral}
\label{App:C}

In this appendix we evaluate the integral, which is necessary to establish what sign in (\ref{16mar25}) corresponds to positive and what to negative energy modules. More precisely 
\begin{equation}
\label{13jun1}
I\equiv\int (z-z_1)^{-\frac{3}{2}}(\bar z-\bar z_1)^{-\frac{3}{2}}\frac{1}{((1+z_1\bar z_1))^\frac{1}{2}}dz_1 d\bar z_1.
\end{equation}
First we convert it to real variables,
\begin{equation}
\label{13jun2}
dz d\bar z=-2i dx dy.
\end{equation}
We then obtain
\begin{equation}
\label{13jun3}
I=-2i \int ((x-x_1)^2 + (y-y_1)^2 )^{-\frac{3}{2}}\frac{1}{((1+x_1^2+y_1^2)^\frac{1}{2}}dx_1dy_1.
\end{equation}
Next, we introduce Schwinger parameters
\begin{equation}
\label{13jun4}
\alpha^{-z}=\frac{1}{\Gamma(z)}\int_0^\infty t^z e^{-\alpha t}\frac{dt}{t}
\end{equation}
to deal with powers. We then get
\begin{equation}
\label{13jun5}
I=-\frac{2i}{\Gamma(\frac{3}{2})\Gamma(\frac{1}{2})}\int dx_1 dy_1 
\int \frac{dt_1}{t_1}\frac{dt_2}{t_2}t_1^{\frac{3}{2}}t_2^{\frac{1}{2}}
e^{-\left[ ((x-x_1)^2+(y-y_1)^2)t_1-(x_1^2+y_1^2+1)t_2\right]}.
\end{equation}
Completing the squares
and
evaluating Gaussian integrals over $x_1$ and $y_1$,
we find
\begin{equation}
\label{13jun8}
I=-\frac{2\pi i}{\Gamma(\frac{3}{2})\Gamma(\frac{1}{2})}
\int \frac{dt_1}{t_1}\frac{dt_2}{t_2}t_1^{\frac{3}{2}}t_2^{\frac{1}{2}}\frac{1}{t_1+t_2}
e^{-\left[ (x^2+y^2)\frac{t_1t_2}{t_1+t_2}+t_2\right]}.
\end{equation}

Next, we change variables
\begin{equation}
\label{13jun9}
\alpha_1=\frac{t_1}{t_1+t_2}, \qquad \alpha_2=t_2,
\end{equation}
\begin{equation}
\label{13jun10}
t_1=\frac{\alpha_1\alpha_2}{1-\alpha_1},\qquad dt_1=d\alpha_1\frac{\alpha_2}{(1-\alpha_1)^2}+\alpha_1\frac{d\alpha_2}{1-\alpha_1}.
\end{equation}
At fixed $t_2$, $\alpha_1$ changes from 0 to 1 when $t_1$ changes from 0 to $\infty$.
Therefore, we have
\begin{equation}
\label{13jun11}
I=-\frac{2\pi i}{\Gamma(\frac{3}{2})\Gamma(\frac{1}{2})}\int_0^1\frac{d\alpha_1}{\alpha_1}\int_0^\infty \frac{d\alpha_2}{\alpha_2}
\alpha_1^{\frac{3}{2}}\alpha_2^1\frac{1}{(1-\alpha_1)^{\frac{3}{2}}}e^{-(x^2+y^2)\alpha_1\alpha_2-\alpha_2}.
\end{equation}
It is straightforward to evaluate the $\alpha_2$ integral with (\ref{13jun4}), which leads to
\begin{equation}
\label{13jun12}
I=-\frac{2\pi i \Gamma(1)}{\Gamma(\frac{3}{2})\Gamma(\frac{1}{2})}\int_0^1d\alpha_1
\alpha_1^{\frac{1}{2}}\frac{1}{(1-\alpha_1)^{\frac{3}{2}}}
\frac{1}{(x^2+y^2)\alpha_1+1}.
\end{equation}

The above integral is divergent. To regularise it, we introduce an extra parameter $m$
\begin{equation}
\label{13jun13}
I_1(m)\equiv\int_0^1d\alpha_1
\alpha_1^{\frac{1}{2}}{(1-\alpha_1)^{m}}
\frac{1}{(x^2+y^2)\alpha_1+1}.
\end{equation}
This integral can be evaluated with Mathematica and gives an expression in terms of Gauss hypergeometric functions. 
By substituting
 $m=-\frac{3}{2}$ we find
\begin{equation}
\label{13jun14}
I_1\left(-\frac{3}{2}\right)=-\frac{\pi}{(1+x^2+y^2)^{\frac{3}{2}}}.
\end{equation}
This is then used to evaluate
\begin{equation}
\label{13jun15}
I=4\pi i\frac{1}{(1+x^2+y^2)^{\frac{3}{2}}}.
\end{equation}

Reinstating the extra factors to get ${\cal O}$ on the left hand side, we find
\begin{equation}
\label{13jun16}
d\frac{i}{2}\int (z-z_1)^{-\frac{3}{2}}(\bar z-\bar z_1)^{-\frac{3}{2}}\frac{1}{(2(1+z_1\bar z_1))^\frac{1}{2}}dz_1 d\bar z_1=
-\frac{4\pi d}{(2(1+z\bar z))^\frac{3}{2}}.
\end{equation}
Then, to reproduce the desired $\varphi^{(-\frac{3}{2},-\frac{3}{2})}$ for the positive energy case (\ref{16mar7}), we need to set
\begin{equation}
\label{13jun17}
d_+=\frac{1}{2\pi AR}.
\end{equation}

\begin{figure}
\centering
\begin{tikzpicture}[scale=1][>=stealth]
\draw  [->] (0,0) -- (0,5.5);
\draw  [very thin] (-1,0) -- (-1,5.5);
\draw  [very thin] (-2,0) -- (-2,5.5);
\draw [->] (-3.5,0) -- (1.5,0) ;
\filldraw 
(0,2) circle (2pt)
(-1,3) circle (2pt)
(-2,2) circle (2pt)
(-1,1) circle (2pt)
;
\draw  [very thin] (0,2) -- (-1,3);
\draw  [very thin] (0,2) -- (-1,1);
\draw  [very thin] (-2,2) -- (-1,3);
\draw  [very thin] (-2,2) -- (-1,1);
\draw(1.5,-0.1) node[anchor=north east,fill=white] {$\bar N$};
\draw(0.1,5.5)  node[anchor=north west,fill=white] {$N$};
\draw(0.1,2)  node[anchor=west,fill=white] {$N_0$};
\draw(0,-0.1)  node[anchor=north,fill=white] {$0$};
\draw(-1,-0.1)  node[anchor=north,fill=white] {$-1$};
\draw(-2,-0.1)  node[anchor=north,fill=white] {$-2$};
\draw(-1,-1) node {a)};
\end{tikzpicture}
$\qquad\qquad$
\begin{tikzpicture}[scale=1][>=stealth]
\draw  [->] (0,0) -- (0,5.5);
\draw  [very thin] (-1,0) -- (-1,5.5);
\draw  [very thin] (-2,0) -- (-2,5.5);
\draw [->] (-3.5,0) -- (1.5,0) ;
\filldraw 
(0,2) circle (2pt)
(-1,3) circle (2pt)
(-2,2) circle (2pt)
(-1,1) circle (2pt)
(0,4) circle (2pt)
(-1,5) circle (2pt)
(-2,4) circle (2pt)
;
\draw  [very thin] (0,2) -- (-1,3);
\draw  [very thin] (0,2) -- (-1,1);
\draw  [very thin] (-2,2) -- (-1,3);
\draw  [very thin] (-2,2) -- (-1,1);
\draw  [very thin] (0,4) -- (-1,5);
\draw  [very thin] (0,4) -- (-1,3);
\draw  [very thin] (-2,4) -- (-1,5);
\draw  [very thin] (-2,4) -- (-1,3);
\draw(1.5,-0.1) node[anchor=north east,fill=white] {$\bar N$};
\draw(0.1,5.5)  node[anchor=north west,fill=white] {$N$};
\draw(0.1,2)  node[anchor=west,fill=white] {$N_0$};
\draw(0,-0.1)  node[anchor=north,fill=white] {$0$};
\draw(-1,-0.1)  node[anchor=north,fill=white] {$-1$};
\draw(-2,-0.1)  node[anchor=north,fill=white] {$-2$};
\draw(-1,-1) node {b)};
\end{tikzpicture}
\caption{a) Non-vanishing $F$ and $G$ are indicated by diagonal lines. These are related by $G(N_0,0)=-F(N_0,-2)=\frac{N_0}{N_0+2}F(N_0-1,-1)=-\frac{N_0}{N_0+2}G(N_0-1,-1)$. $N_0$ can take any integer value except $0$ and $-2$. One can ''join'' any number of such modules as shown on b) for two modules.}
\label{fig:7}
\end{figure}
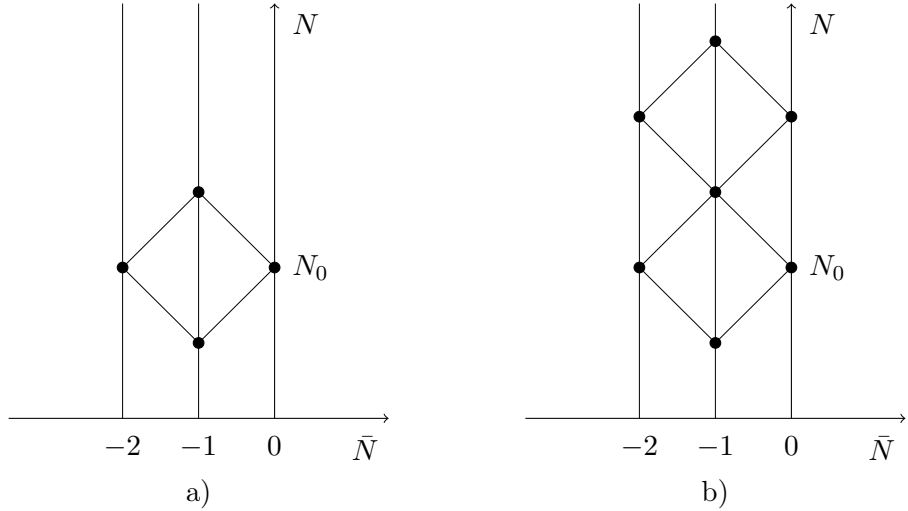

\begin{figure}
\centering
\begin{tikzpicture}[scale=1][>=stealth]
\draw  [->] (0,-2.5) -- (0,1);
\draw  [very thin] (-1,-2.5) -- (-1,0.5);
\draw  [very thin] (-2,-2.5) -- (-2,0.5);
\draw [->] (-2.5,0) -- (1,0) ;
\draw  [very thin] (-2.5,-1) -- (0.5,-1);
\draw  [very thin] (-2.5,-2) -- (0.5,-2);
\filldraw 
(0,0) circle (2pt)
(-1,-1) circle (2pt)
(-2,-2) circle (2pt)
(-2,0) circle (2pt)
(0,-2) circle (2pt)
;
\draw  [very thin] (-2,-2) -- (0,0);
\draw  [very thin] (-2,0) -- (0,-2);
\draw(1,-0.1) node[anchor=north east,fill=white] {$\bar N$};
\draw(0.1,1)  node[anchor=north west,fill=white] {$N$};
\draw(-0.1,0.1)  node[anchor=south east,fill=white] {$0$};
\draw(-1,0.1)  node[anchor=south,fill=white] {$-1$};
\draw(-2,0.1)  node[anchor=south,fill=white] {$-2$};
\draw(0.1,-1)  node[anchor=west,fill=white] {$-1$};
\draw(0.1,-2)  node[anchor=west,fill=white] {$-2$};
\end{tikzpicture}
\caption{The only condition on the non-vanishing coefficients is $F(-2,-2)+F(-1,-1)+G(-2,0)+G(-1,-1)=0$.}
\label{fig:8}
\end{figure}
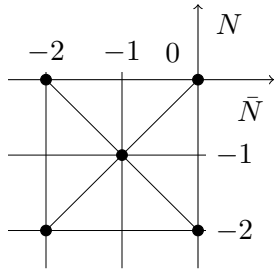

\section{Short representations of $iso(3,1)$ with non-trivial coefficient functions}
\label{App:flat}

Away from special points, the general solution to the flat space consistency conditions (\ref{4may10}) is given by (\ref{27mar2}).
As for the $so(3,2)$ case, at special points the consistency conditions degenerate and, hence, have to be studied separately. We analysed  the flat space consistency conditions at these points. A complete set of solutions of these conditions that involve non-vanishing $F$ and $G$ at a finite set of weights in the weight space is presented on  Fig \ref{fig:7} and Fig \ref{fig:8}. By setting $A$, $B$, $C$ and $D$ to zero at all other points, we obtain short Poincare modules with non-trivially realised translations. Analysis of properties of these modules we leave for future research.

\bibliography{shortrep}
\bibliographystyle{JHEP}

\end{document}